\documentclass[article]{JHEP}
\usepackage{amsmath,amssymb}

\newcommand{\be}{\begin{equation}}
\newcommand{\ee}{\end{equation}}
\newcommand{\bea}{\begin{eqnarray}}
\newcommand{\eea}{\end{eqnarray}}
\newcommand{\bean}{\begin{eqnarray*}}
\newcommand{\eean}{\end{eqnarray*}}

\def\beq{\begin{equation}}

\def\eeq{\end{equation}}

\relax

\preprint{HUTP--01/A061\\  {\tt hep-th/0112231}}

\title{Towards Vacuum Superstring Field Theory:\\ The Supersliver}

\author{Marcos Mari\~no and Ricardo Schiappa
\\
Department of Physics,\\
Harvard University,\\ Cambridge, MA 02138, USA\\
\email{marcos, ricardo@lorentz.harvard.edu}
}

\abstract{We extend some aspects of vacuum string field theory to 
superstring field theory in Berkovits' formulation, and we study the 
star algebra in the fermionic matter sector. After clarifying the 
structure of the interaction vertex in the operator formalism of Gross 
and Jevicki, we provide an algebraic construction of the supersliver state 
in terms of infinite--dimensional matrices. This state is an idempotent 
string field and solves the matter part of the equation of motion of 
superstring field theory 
with a pure ghost BRST operator. We determine the spectrum of eigenvalues 
and eigenvectors of the infinite--dimensional matrices of Neumann 
coefficients in the fermionic matter sector. We then analyze coherent states 
based on the supersliver and use them in order to construct higher--rank 
projector solutions, as well as to construct closed subalgebras of the star 
algebra in the fermionic matter sector. Finally, we show that the geometric 
supersliver is a solution to the superstring field theory equations of 
motion, including the (super)ghost sector, with the canonical choice 
of vacuum BRST operator recently proposed by Gaiotto, Rastelli, Sen 
and Zwiebach.
}


\begin{document}



\vfill

\eject

\section{Introduction and Summary} \label{s0}

In the last two years, the search for nonperturbative information in string 
field theory \cite{WITTENSFT, Kostelecky-Samuel} has experienced a renewed 
interest mainly due to a series of conjectures by Sen \cite{Sen-1, 
Sen-2, Sen-3} (also see \cite{sen} for a review and a list of references). 
These conjectures have been tested numerically to a high degree of precision 
in level truncated cubic string field theory, and some of them 
have been proven in boundary string field theory (see, \textit{e.g.}, 
\cite{ohmoreview} for a review and a list of references). In the 
meantime, the elegant construction of Berkovits \cite{berko, berko2, 
berkoreview, berkofour} has emerged as a promising candidate for an open 
superstring field theory describing the NS sector: in here, Sen's conjectures 
about the fate of the tachyon in the non--BPS $D9$--brane have been 
successfully tested by level truncation to a high level of accuracy 
\cite{berkotachyon, bsz, desmet, in}, and kink solutions have been found 
that describe lower--dimensional $D$--branes \cite{kink} (see, 
\textit{e.g.}, \cite{thesis} for a review and a more complete list of 
references).

So far, most of our understanding about tachyon condensation in both cubic 
string field theory and Berkovits' superstring field theory is based 
on level--truncated computations and it would be of course desirable to 
have an analytical control over the problem. For the bosonic string, Rastelli, 
Sen and Zwiebach have proposed in a series of papers \cite{RSZ, 
RSZclassical, RSZhalf, RSZboundary, RSZvacuum} a new approach to this problem 
called vacuum string field theory (VSFT). In VSFT, the form of the cubic 
string field theory action around the tachyonic vacuum is postulated by 
exploiting some of the expected properties it should have (like the absence 
of open string states). Then one can show that this theory has solutions 
that describe the perturbative vacuum and the various $D$--branes. 
In particular, the matter sector of the maximal $D25$--brane is described by 
a special state called the sliver. This state was first constructed 
geometrically by Rastelli and Zwiebach \cite{RZ} and then algebraically by 
Kostelecky and Potting \cite{KP}, and it is an idempotent state of the string 
field star algebra, in the matter sector. The construction of VSFT has been 
recently completed in \cite{GRSZ}, where Gaiotto, Rastelli, Sen and Zwiebach 
have proposed a canonical choice of the ghost BRST operator around the 
vacuum, with which they identified closed string states; and also in 
\cite{RSZspectro}, where Rastelli, Sen and Zwiebach have found the eigenvalue 
and eigenvector spectrum of the Neumann matrices, which could allow for a 
proper definition of the string field space. The study of VSFT has also unveiled 
beautiful algebraic structures in cubic string field theory (for example, 
projectors of arbitrary rank in the star algebra have been constructed in 
detail in \cite{RSZhalf, GT, GT2}).

The main purpose of this paper is to give the first steps towards the
construction of vacuum superstring field theory around the tachyonic vacua
of the non--BPS maximal $D9$--brane in Type IIA superstring theory, and to 
explore the algebraic structure of the star algebra in the fermionic part of 
the matter sector. In section 2, we begin with a review of Berkovits' open 
superstring field theory for the NS sector and discuss the general features of 
vacuum superstring field theory. We shall show in detail that, assuming a pure 
ghost BRST operator around the vacuum as in VSFT, Berkovits' equation of motion 
for the superstring field admits factorized solutions whose matter part is an 
idempotent state of the star algebra. In a sense, idempotency is even more 
useful in Berkovits' theory since it drastically reduces the nonlinearity of 
the equation of motion. Idempotent string field solutions can be 
constructed in the GSO(+) sector or in both GSO($\pm$) sectors. 

In order to construct idempotent states in superstring field theory, 
one first has to understand in detail the structure of the star algebra in the 
fermionic matter sector. To do that, we use the operator construction of 
the interaction vertex for the superstring due to Gross and Jevicki
\cite{GJ3}, which extends their previous work on the bosonic string 
\cite{GJ, GJ2} to the NSR superstring. In section 3 we review some of the 
relevant results and we further clarify the structure of the vertex. This 
allows us to write the Neumann coefficients in terms of two simple 
infinite--dimensional matrices which shall play a key r\^ole in the 
constructions of this paper.  

Given any boundary conformal field theory (BCFT) one can construct 
geometrically a special state which is an idempotent of the star algebra
\cite{RSZboundary}. When the BCFT is that of a $D25$--brane, this state is 
called the sliver\footnote{Strictly speaking, the construction of the sliver 
state is purely geometric and is thus valid for arbitrary BCFT's. However, in 
this paper, we shall use the denotation of ``sliver'' for the particular BCFT 
associated to the maximal brane in flat space.} \cite{RZ, RSZ}. This geometric 
construction extends in a very natural way to the BCFT given by the NS sector 
of the open superstring which describes the unstable $D9$--brane. This yields 
an idempotent state that we call the supersliver. The matter part of the 
supersliver is a product of two squeezed states: one made of bosonic 
oscillators (the bosonic sliver previously considered in \cite{KP, 
RSZclassical}) and the other made of fermionic oscillators, that we 
shall call the fermionic sliver. Although the geometric construction 
gives a precise determination of the fermionic sliver, it is important 
for many purposes to have an algebraic construction as well. 
In section 4, and making use the results of section 3 for the 
interaction vertex, we find a simple expression for the fermionic 
sliver in terms of infinite--dimensional matrices, as in \cite{KP, 
RSZclassical}, and we compare the result to the geometric construction. 
We also briefly address the supersliver conservation laws. In section 5 we 
use the techniques recently introduced in \cite{RSZspectro}
to determine the eigenvalue spectrum and the eigenvectors of
the various infinite--dimensional matrices involved in the fermionic star 
algebra, including the matrices of Neumann coefficients.

Once the fermionic sliver has been constructed algebraically, one can take 
it as a sort of ``vacuum state'' in order to build fermionic coherent
states. This we do in section 6, where after constructing these fermionic 
coherent states on the fermionic sliver, we study their star algebras. 
As in \cite{RSZhalf}, one can use these coherent states to construct 
higher--rank projectors of the fermionic star algebra. We shall show 
that one can also construct closed fermionic star subalgebras. These 
star subalgebras provide new idempotent states which yield new 
solutions to the vacuum superstring field theory equation of motion. However, 
some of them turn out to be related to the fermionic sliver by gauge transformations. 

In section 7 we consider the ghost/superghost sector, and we show that if
one chooses the vacuum BRST operator to 
be the recent canonical choice of Gaiotto, Rastelli, Sen and Zwiebach, 
\cite{GRSZ}, then the geometrical sliver is a solution to Berkovits'
superstring field theory equations of motion, \textit{i.e.}, we solve 
the equations of motion in the (super)ghost sector.

Finally, in section 8 we state some conclusions and open problems for 
the future. In the Appendix we give some of the details needed 
in the proof that the structure of the vertex found in section 3 agrees 
with the explicit expressions found by Gross and Jevicki in \cite{GJ3} 
using conformal mapping techniques.

\section{Berkovits' Superstring Field Theory}

\subsection{A Short Review of Berkovits' Superstring Field Theory}

In this paper, we shall study the non--GSO projected open superstring in 
the NS sector. In the matter sector, there are two fermions 
$\psi^{\pm}(\sigma)$ with the mode expansion

\begin{equation}
\label{fexpansion}
\psi_{\pm}^{\mu}(\sigma) =\sum_{r \in {\bf Z} +{1\over2}}
{\rm e}^{\pm ir \sigma}\psi^{\mu}_r,
\end{equation}

\noindent
where the modes satisfy the anticommutation relations

\begin{equation}
\label{modes}
\{ \psi_r^{\mu}, \psi^{\nu}_s \} =\eta^{\mu \nu}\delta_{r+s,0}.
\end{equation}

\noindent
We will therefore write $\psi^{\dagger}_r =\psi_{-r}$ for $r>0$. The
ghost/superghost sector includes the $b,c$, and the $\beta,\gamma$, system 
and we bosonize the last one in the standard way \cite{fms}:

\begin{equation}
\beta=\partial \xi {\rm e}^{-\phi}, \,\,\,\,\,\,\,\,\, 
\gamma =\eta {\rm e}^{-\phi}.
\end{equation}

A superstring field theory describing the GSO--projected NS sector of the 
open superstring was proposed by Berkovits in \cite{berko} (recent 
reviews can be found in, \textit{e.g.}, \cite{berkoreview, thesis}). In 
this theory, the string field $\Phi$ is Grassmann even, has zero ghost 
number and zero picture number. The action has the structure of a WZW model:

\begin{equation}
\label{berkoaction}
S[\Phi]={1 \over 2} \int \biggl( ({\rm e}^{-\Phi} Q_B {\rm e}^{\Phi})
({\rm e}^{-\Phi} \eta_0 {\rm e}^{\Phi})
-\int_0^1 dt ({\rm e}^{-t\Phi} \partial_t {\rm e}^{t\Phi})
\{ ({\rm e}^{-t\Phi} Q_B {\rm e}^{t\Phi}),
({\rm e}^{-t\Phi} \eta_0 {\rm e}^{t\Phi}) \} \biggr),
\end{equation}

\noindent
where $Q_{B}$ is the BRST operator of the superstring and $\eta_{0}$ 
the zero--mode of $\eta$ (the bosonized superconformal ghost) \cite{fms}. In a 
WZW interpretation of this model, these operators play the role of a 
holomorphic and an anti--holomorphic derivatives, respectively. In this
action, the integral and the star products are evaluated with Witten's 
string field theory interaction \cite{WITTENSFT}. The exponentiation of the 
string field $\Phi$ is defined by a series expansion with star products: ${\rm
e}^{\Phi}={\cal I}+ \Phi+ {1 \over 2} \Phi \star \Phi + \cdots$, where
${\cal I}$ is the identity string field. As usual, we refer to the first term
in (\ref{berkoaction}) as the kinetic term and to the second one as the
Wess--Zumino term. It can be shown that the equation of motion derived 
from this action is \cite{berko}:

\begin{equation}
\label{ssftmotion}
\eta_0 \Bigl( {\rm e}^{-\Phi} Q_B {\rm e}^{\Phi}\Bigr)=0.
\end{equation}

\noindent
The action (\ref{berkoaction}) has a gauge symmetry given by

\begin{equation}
\label{gaugesym}
\delta {\rm e}^{\Phi}=\Xi_L {\rm e}^{\Phi} + {\rm e}^{\Phi} \Xi_R,
\end{equation}

\noindent
where the gauge parameters $\Xi_{L,R}$ satisfy

\begin{equation}
Q_B \Xi_L =0, \,\,\,\,\,\,\,\,\, \eta_0 \Xi_R =0.
\end{equation}

One can include GSO$(-)$ states by introducing Chan--Paton--like degrees of
freedom \cite{berkotachyon, bsz}. The string field then reads,

\begin{equation}
\label{fisectors}
\Phi= \Phi_+ \otimes {\bf 1} + \Phi_- \otimes \sigma_1,
\end{equation}

\noindent
where $\Phi_{\pm}$ are respectively in the GSO$(\pm)$ sectors, and
$\sigma_1$ is one of the Pauli matrices. The $Q_B$ and $\eta_0$ operators
also have to be tensored with the appropriate matrices:

\begin{equation}
{\hat Q}_B=Q_B \otimes \sigma_3, \,\,\,\,\,\,\,\,\, {\hat \eta}_0 = \eta_0 
\otimes \sigma_3.
\end{equation}

\noindent
The action is again given by one--half times (\ref{berkoaction}), where the 
bracket now includes a trace over the Chan--Paton--like matrices (the $1/2$ 
factor is included to compensate for the trace over the matrices).
The gauge symmetry is given again by (\ref{gaugesym}), where $\Xi_{L,R}$
take values in both sectors as in (\ref{fisectors}). It has been shown
that Berkovits superstring field theory correctly reproduces the four--point
tree amplitude in \cite{berkofour}, and it can be used to computed the NS 
tachyon potential in level truncation (see \cite{thesis} for a review), giving
results which are compatible with Sen's conjectures.

\subsection{Superstring Field Theory Around a Classical Solution}

In the cubic theory of Witten, one can consider a particular solution of
the classical equations of motion, $\Phi_0$, and study fluctuations around it:
$\Phi=\Phi_0 + {\tilde \Phi}$. It is easy to see that the action governing
the fluctuations $\tilde \Phi$ has the structure of the original action for
$\Phi$, but with a different BRST operator, ${\cal Q}$. Bosonic VSFT, as 
formulated in the series of papers \cite{RSZ, RSZclassical, RSZhalf, 
RSZboundary, RSZvacuum}, is based on two assumptions:

\bigskip

\noindent
\textbf{1)} First, it is assumed that, when one expands around the tachyonic 
vacuum, the new BRST operator ${\cal Q}$ has vanishing cohomology and is 
made purely of ghost operators.

\bigskip

\noindent
\textbf{2)} Second, it is assumed that all $Dp$--brane solutions of VSFT have 
the factorized form:

\begin{equation}
\label{factorized}
\Phi = \Phi_{g} \otimes \Phi_m,
\end{equation}

\noindent
where $\Phi_{g,m}$ denote states containing only ghost and only matter 
modes, respectively. Since the star product factorizes into the ghost and 
the matter sector, and since we have assumed that ${\cal Q}$ is pure ghost, 
the equations of motion split into:

\begin{equation}
{\cal Q} \Phi_g + \Phi_g \star\Phi_g =0,
\end{equation}

\noindent
and

\begin{equation}
\label{bidem}
\Phi_m \star \Phi_m =\Phi_m.
\end{equation}

\noindent
The second equation says that the matter part is an idempotent of the
star algebra (where the star product is now restricted to the matter 
sector). If these assumptions hold, the string field action evaluated 
at a solution of the form (\ref{factorized}) is simply proportional to 
the BPZ norm of $|\Phi_m\rangle$, and this allows one to compare in a simple 
way ratios of tensions of different $D$--branes \cite{RSZclassical, RSZboundary}.

An interesting question is to which extent are these assumptions valid in
Berkovits' superstring field theory. In order to answer this question, 
the first step is to analyze the fluctuations around a solution to the 
equations of motion. This was first addressed by Kluson in \cite{Kluson}, 
where it was shown that with an appropriate parameterization 
of the fluctuations, the equation of motion is identical to (\ref{ssftmotion}), 
albeit with a deformed $Q$ operator. It was thus concluded (without 
proof) in \cite{Kluson} that the action for the fluctuation should have the 
form (\ref{berkoaction}) with the deformed operator. We shall now derive the 
equation of motion in a slightly different way from the one presented in 
\cite{Kluson}, and this will allow us to show that the action is indeed of the 
required form by direct computation.

Let us define $G={\rm e}^{\Phi}$, the exponential of the string field
that appears in Berkovits' action. Let $\Phi_0$ be a solution to the
classical equations of motion (\ref{ssftmotion}) and let us consider
a fluctuation around this solution parameterized as follows 
\cite{Kluson},

\begin{equation}
\label{fluctu}
G=G_0 \star h, \,\,\,\,\,\,\,\,\, G_0={\rm e}^{\Phi_0}, 
\,\,\,\,\,\,\,\,\, h={\rm e}^{\phi}.
\end{equation}

\noindent
Since Berkovits' action has the structure of a WZW theory, one should
expect an analog of the Polyakov--Wiegmann equation \cite{PW} to be valid.
In fact, it is easy to show (by using for example the geometric formulation
of \cite{Kluson, klu2}) that the action (\ref{berkoaction}) satisfies

\begin{equation}
\label{pw}
S[G_0\star h]=S[G_0] + S[h] -\int (G_0^{-1} Q_{B} G_0) ( h \eta_0 h^{-1}),
\end{equation}

\noindent
for arbitrary $G_0$ and $h$. The effective action for the fluctuations is
then

\begin{equation}
\label{effec}
S_{\rm eff}[h] = S[h]-\int (G_0^{-1} Q_{B} G_0) ( h \eta_0 h^{-1}).
\end{equation}

\noindent
Let us now obtain the equation of motion satisfied by $h$. Varying 
$S[h]$, one obtains

$$
\int h^{-1} \delta h \eta_0 (h^{-1} Q_{B} h),
$$

\noindent
and from the extra term in $S_{\rm eff}[h]$ one gets

$$
\int h^{-1} \delta h \eta_0 (h^{-1} A h),
$$

\noindent
where we denoted $A=G^{-1}_0 Q_{B} G_0$ and we have used the equation
of motion $\eta_0(A)=0$. Putting both pieces together, one finds that

\begin{equation}
\eta_0 ( h^{-1} Q_{B} h + h^{-1}A h -A)=0.
\end{equation}

\noindent
Therefore, the equation of motion is identical to (\ref{ssftmotion}) but 
with the deformed $Q$ operator:

\begin{equation}
\label{deform}
Q_A(X) = Q_{B}(X) + AX -(-1)^X XA.
\end{equation}

\noindent
One can moreover easily prove \cite{Kluson} that the new operator satisfies 
all the axioms of superstring field theory (it is a nilpotent derivation and 
it anticommutes with $\eta_0$).

We shall now show that $S_{\rm eff}[h]$ has in fact the structure of 
(\ref{berkoaction}) but with the operator $Q_A$. For that, we simply 
need to notice that

\begin{equation}
\label{calcul}
\int A ( h \eta_0 h^{-1}) = {1 \over 2} \int \biggl(
(h^{-1}A h -A) (h^{-1} \eta_0 h)  - \int_0^1 dt A
\partial_t(\hat h \eta_0 \hat h^{-1} - \hat
h^{-1} \eta_0 \hat h)\biggr),
\end{equation}

\noindent
where we have used integration by parts with respect to $\eta_0$, and the fact 
that $\Phi_0$ satisfies its equation of motion. We have also denoted
$\hat h ={\rm e}^{t\phi}$. The first term in the RHS of (\ref{calcul})
when added to the kinetic term in $S[h]$ gives a kinetic term with the
$Q_A$ operator, while the second term when added to the
Wess--Zumino term in $S[h]$ gives a Wess--Zumino term with $Q_A$.
The conclusion of this computation is that the action for the
fluctuations is simply

\begin{equation}
S_{\rm eff}[h]= {1 \over 2} \int \biggl( ({\rm e}^{-\phi} Q_A {\rm e}^{\phi})
({\rm e}^{-\phi} \eta_0 {\rm e}^{\phi})
-\int_0^1 dt ({\rm e}^{-t\phi} \partial_t {\rm e}^{t\phi})
\{ ({\rm e}^{-t\phi} Q_A {\rm e}^{t\phi}),
({\rm e}^{-t\phi} \eta_0 {\rm e}^{t\phi}) \} \biggr),
\end{equation}

\noindent
as anticipated in \cite{Kluson}.

Let us now consider the superstring field theory describing the non--BPS 
$D9$--brane, {\it i.e.}, Berkovits' superstring field theory including 
both the GSO($\pm$) sectors. It has been shown in level truncation that this 
theory has two symmetric vacua where the tachyon condenses. According 
to Sen's conjectures, at any of these two vacua there are no open superstring 
degrees of freedom. Let us then choose one of these vacua and study the action 
for fluctuations around it. As we have seen, the action for the fluctuations 
has the same form as the original one, but with a different BRST operator, that 
we shall now denote by ${\cal Q}$. According to Sen's conjectures, at this 
chosen vacuum there are no open string degrees of freedom and it is 
thus natural to assume, as in the VSFT for the bosonic string, that the new 
BRST operator has vanishing cohomology and is made purely of (super)ghost 
operators. In addition we will also assume that this operator annihilates 
the identity,

\begin{equation}
{\cal Q} {\cal I}=0.
\end{equation}

\noindent
This condition, although very natural, is strictly not necessary in 
order to preserve some of the basic features of bosonic VSFT. In the 
superstring case however, it is crucial. It was noticed in \cite{RSZ} that 
operators of the form ${\cal Q}=c_0 + \sum_n u_n c_n$ also have vanishing 
cohomology in the superstring case. In particular, the ${\cal Q}$ operator 
recently proposed in \cite{GRSZ} for the bosonic VSFT is of this form and 
annihilates the identity after some proper regularization, so that in principle 
it is a possible candidate for the superstring as well (where the 
superconformal ghost sector would be handled separately). We shall 
come back to this question in section 7.

With these assumptions at hand, and given the fact that the action around the
vacuum has the same form as the original one but with a pure (super)ghost 
operator ${\cal Q}$, it is now easy to show that the ansatz (\ref{factorized}) 
solves the superstring field theory equations of motion if $\Phi_m$ is 
idempotent and $\Phi_g$ satisfies

\begin{equation}
\label{ghosteq}
\eta_0 \Bigl( {\rm e}^{-\Phi_g} {\cal Q} {\rm e}^{\Phi_g} \Bigr)=0.
\end{equation}

\noindent
In order to see this, notice that idempotency of $\Phi_m$ and factorization of
the star product in matter and ghost parts yields

\begin{equation}
\label{exfi}
{\rm e}^{\Phi}= {\rm e}^{\Phi_g} \otimes \Phi_m + {\cal I} -\Phi_m,
\end{equation}

\noindent
and, since ${\cal Q}$ kills the identity and is pure ghost, one has

\begin{equation}
{\cal Q} {\rm e}^{\Phi}=\bigl( {\cal Q} {\rm e}^{\Phi_g} \bigr)
 \otimes \Phi_m.\end{equation}

\noindent
Using again idempotency of $\Phi_m$, the equation of motion becomes:

\begin{equation}
\biggl( \eta_0 \Bigl( {\rm e}^{-\Phi_g} {\cal Q} {\rm e}^{\Phi_g} \Bigr)
\biggr)\otimes \Phi_m=0.
\end{equation}

\noindent
Therefore, the above conditions are sufficient to solve the equations of
motion. In the same way, one can show that in these circumstances the
action factorizes as

\begin{equation}
S= K \langle \Phi_m | \Phi_m \rangle,
\end{equation}

\noindent
where

\begin{equation}
K=S[\Phi_g].
\end{equation}

Let us now look at the gauge symmetry of the new action around the tachyon
vacuum. We are particularly interested in transformations that preserve the 
structure of (\ref{exfi}). Since both ${\cal Q}$ and $\eta_0$ annihilate the 
identity, it is easy to see that the gauge transformation 
(\ref{gaugesym}) with

\begin{equation}
\Xi_L=\Xi_m \otimes {\cal I}_g, \,\,\,\,\,\,\,\,\, \Xi_R=-\Xi_m \otimes 
{\cal I}_g,
\end{equation}

\noindent
preserves (\ref{exfi}). This gauge transformation leaves $\Phi_g$ invariant 
and changes $\Phi_m$ as follows:

\begin{equation}
\label{gaugem}
\delta \Phi_m = [\Xi_m , \Phi_m]_\star,
\end{equation}

\noindent
where $[A, B]_\star=A\star B -B \star A$ is the commutator 
in the star algebra. 
Notice that this transformation preserves idempotency of $\Phi_m$ at linear
order. The gauge symmetry (\ref{gaugem}) is precisely the one that appears
in bosonic VSFT when ${\cal Q}$ annihilates the identity 
\cite{RSZclassical, RSZboundary,GRSZ}.

The condition of idempotency of $\Phi_m$ in the non--GSO projected theory
involves in fact two different conditions. In general, a matter
string field $\Phi_m$ has components in both GSO($\pm$) sectors,

\begin{equation}
\Phi_m=\Phi_m^{+} \otimes {\bf 1} + \Phi_m^{-} \otimes \sigma_1.
\end{equation}

\noindent
In this equation, $\Phi_m^{\pm}$ is Grassmann even (odd),
and idempotency of $\Phi_m$ is equivalent to the following equations

\begin{eqnarray}
\label{comproj}
\Phi_m^+ \star \Phi_m^+ + \Phi_m^- \star \Phi_m^-&=&\Phi_m^+,\nonumber\\
\Phi_m^+ \star \Phi_m^- + \Phi_m^- \star \Phi_m^+&=&\Phi_m^-.
\end{eqnarray}

\noindent

One particular solution is of course to take $\Phi_m^+$ as an idempotent 
state and $\Phi_m^-=0$. The matter supersliver state that we will
discuss later is an example of such a solution. Another possibility is
to take $\Phi_m^+$ an idempotent and $\Phi_m^-$ a nilpotent state satisfying
the second equation in (\ref{comproj}). In section 6 we will construct
solutions with these characteristics, although we will also show that 
they are related to the supersliver solution by gauge 
transformations at the vacuum.

\section{Neumann Coefficients and Overlap Equations}

In this section we review some of the results of \cite{GJ3} and
we explain in detail the structure of the overlap equations
involving the matter part of the fermionic sector.

\subsection{The Identity}

As in bosonic string field theory, the simplest vertex in superstring field 
theory is the integration, which corresponds to folding the string and 
identifying the two halves \cite{WITTENSFT} thus defining the identity string 
field $|I \rangle$,

\begin{equation}
\int \Phi = \langle I | \Phi \rangle.
\end{equation}

\noindent
In the bosonic case, the overlap condition defining the identity is simply 
$x(\pi -\sigma)=x(\sigma)$. In the fermionic case, and due to the conformal 
weight $h=1/2$, the precise conditions are as follows:

\begin{eqnarray}\label{threetwo}
\left( \psi_+ (\sigma)-i\psi_+ (\pi-\sigma) \right) |I\rangle &=&0, 
\,\,\,\,\,\,\,\,\, 0\le \sigma \le {\pi \over 2}, \nonumber\\
\left( \psi_- (\sigma)+i\psi_- (\pi-\sigma) \right) |I\rangle &=&0, 
\,\,\,\,\,\,\,\,\, 0\le \sigma \le {\pi \over 2}.
\end{eqnarray}

\noindent
The different sign in the second equation is due to the NS boundary
conditions $\psi_-(0)=\psi_+(0)$, $\psi_-(\pi)=-\psi_+(\pi)$. As usual, we
can define a single antiperiodic fermion field $\psi(\sigma)$ in the interval 
$[-\pi,\pi]$ by declaring that $\psi(\sigma)=\psi_+ (\sigma)$ for $0 \le \sigma 
\le \pi$, and $\psi(\sigma)=\psi_- (-\sigma)$ for $-\pi \le \sigma \le 0$. In 
terms of this single field, the overlap conditions (\ref{threetwo}) read

\begin{equation}
\label{twoverlap}
\psi (\sigma) = \begin{cases}
i\psi (\pi -\sigma),& 0\le |\sigma| \le {\pi \over 2},  \\
 -i\psi (\pi -\sigma),& {\pi \over 2}\le |\sigma| \le \pi.
\end{cases}
\end{equation}

\noindent
This condition leads to the following relation for the modes

\begin{equation}
\label{mmodes}
\left(
\begin{array}{c}
\psi_r \\
\psi_{-r}\end{array}\right)=
 \left(
\begin{array}{cc}
 M_{rs} & {\widetilde M}_{rs} \\
-{\widetilde M}_{rs}& -M_{rs}\end{array}\right)\left(
\begin{array}{c}
\psi_r \\
\psi_{-r}\end{array}\right),
\end{equation}

\noindent
where the matrices $M$, $\widetilde M$, are defined by

\begin{eqnarray}
\label{mm}
M_{rs}&=&-{2 \over \pi} {i^{r-s}\over r+s}, \,\,\,\,\,\,\,\,\,
\quad r=s \,\,\,\, {\rm (mod \, 2)},\\
{\widetilde M}_{rs} &=& {2 \over \pi} {i^{r+s}\over s-r},
\,\,\,\,\,\,\,\,\, \quad r=s+1 \,\,\,\, {\rm (mod \,2)}.
\end{eqnarray}

\noindent
These matrices will play an important role in this paper. They satisfy the
following properties:

\begin{eqnarray}
M^2 -{\widetilde M}^2 = 1, & \quad [M,\widetilde M]=0,\\
{\overline M}=M^T=M, & \quad {\overline {\widetilde M}}=
-{\widetilde M}^T={\widetilde M}.
\end{eqnarray}

\noindent
From (\ref{mmodes}) one obtains the following relation between positive and
negative modes for the fermion fields that annihilate the identity,

\begin{equation}
\label{mmmodes}
\psi_r =\Bigl( {\widetilde M \over 1-M} \Bigr)_{rs} \psi_{-s} .
\end{equation}

\noindent
Using this relation, one can then show that the identity is a squeezed 
state,

\begin{equation}
\label{ident}
| I \rangle = {\cal N}_I \exp\Bigl[ {1 \over 2} \eta_{\mu \nu} \sum_{r,s\ge 
1/2} \psi_{-r}^{\mu} I_{rs} \psi_{-s}^{\nu} \Bigr]|0\rangle,
\end{equation}

\noindent
where

\begin{equation}
\label{idenm}
I={ {\widetilde M} \over 1-M}.
\end{equation}

\noindent
This equation can be obtained acting with $\psi_r^{\mu}$ on $|I\rangle$ and 
using (\ref{mmmodes}). In (\ref{ident}), ${\cal N}_I$ is a normalization
constant that we shall determine later, when we discuss the supersliver.
One can also determine the coefficients $I_{rs}$ explicitly by using conformal
mapping techniques. The result, derived in \cite{GJ3}, is the following. 
Defining the coefficients

\noindent
\begin{equation}
\hat u_{2n}=\hat u_{2n+1} = \left( \begin{array}{c} -1/2 \\
n\end{array}\right)= {(-1)^n (2n-1)!! \over 2^n n!},
\end{equation}

\noindent
one has

\begin{equation}
\label{expli}
I_{rs}=i^{r+s} \biggl( {I_{nm}^+ \over r+s}
-{I^-_{nm}\over r-s} \biggr),\,\,\,\,\,  r=n+1/2, \,\,\ s=m+1/2,
\end{equation}

\noindent
where

\begin{equation}
\label{idecoeffs}
I_{nm}^{\pm} = \begin{cases}
-m\hat u_n \hat u_m,& n={\rm even}, \, m={\rm odd},  \\
\pm n \hat u_n \hat u_m,& n={\rm odd}, \, m={\rm even}, \\
0, & {\rm otherwise}.
\end{cases}
\end{equation}

\noindent
One can check that this explicit expression satisfies the equation
(\ref{idenm}) (see the Appendix).

\subsection{Interaction Vertex and Overlap Equations}

The interaction vertex, $|V_{3}\rangle$, involves the gluing of three strings 
and determines the star algebra multiplication rule,

\begin{equation}
| \Phi \star \Psi \rangle_{(3)} = {}_{(1)} \langle \Phi | 
{}_{(2)} \langle \Psi | | V_{3} \rangle_{(123)}.
\end{equation}

\noindent
In the operator formulation, this vertex involves a set of
infinite--dimensional matrices whose entries are called Neumann
coefficients. Usually, in order to find an explicit expression for these
coefficients, one uses conformal mapping techniques. On the other hand, 
in order to understand the structural properties of these matrices, it 
turns out to be very convenient to analyze the overlap equations as 
well. In this section we shall deduce an expression for the Neumann 
coefficients in terms of the matrices $M$, $\widetilde M$, which will 
be very useful in the following. The starting point is the overlap equation 
for the three string interaction vertex. This overlap equation simply states 
that the interaction is obtained by gluing the halves of the three strings in 
the usual way \cite{WITTENSFT}. In the fermionic case the equation reads
\cite{GJ3}:

\begin{equation}
\label{overlap}
\Big( \psi^a (\sigma) - i \psi^{a-1}(\pi -\sigma) \Big) | V_{3} 
\rangle =0, \,\,\,\,\,\,\,\,\, 0\le \sigma \le {\pi \over 2}, 
\,\,\,\,\,\,\,\,\, a=1,2,3.
\end{equation}

\noindent
The index $a$ labels each of the three strings. As in \cite{GJ}, it is
convenient to diagonalize this condition by introducing the following
discrete Fourier transforms

\begin{eqnarray}
q& =& {1 \over {\sqrt 3}} \left( \psi^1 + \omega \psi^2 + \bar \omega 
\psi^3 \right),\\
q_3&=& {1 \over {\sqrt 3}} \left( \psi^1 + \psi^2 + \psi^3 \right),
\end{eqnarray}

\noindent
together with their adjoints,

\bea
\bar q^{\dagger}& =& {1 \over {\sqrt 3}} \left( (\psi^1)^{\dagger} + \omega
(\psi^2)^{\dagger} + \bar \omega (\psi^3)^{\dagger} \right),\\
q^{\dagger}_3&=& {1 \over {\sqrt 3}} \left( (\psi^1)^{\dagger} +
(\psi^2)^{\dagger} + (\psi^3)^{\dagger} \right),
\eea

\noindent
where $\omega={\rm e}^{2\pi i \over 3}$ is a cubic root of unity.
The overlap conditions give the following condition for $q_3$,

\begin{equation}
\label{qthreecond}
q_3(\sigma)=i q_3 (\pi -\sigma), \,\,\,\,\,\,\,\,\, 0\le \sigma
\le {\pi \over 2},
\end{equation}

\noindent
which is identical in structure to (\ref{twoverlap}). On the other
hand, for $q(\sigma)$ we find

\begin{equation}
\label{compoverlap}
q(\sigma) = \begin{cases}
i\omega q (\pi -\sigma),& 0\le \sigma \le {\pi \over 2},  \\
 -i \bar \omega q (\pi -\sigma),& {\pi \over 2}\le \sigma \le \pi.
\end{cases}
\end{equation}

\noindent
The overlap conditions for $q$ then yield the following relation between 
the modes,

\begin{equation}
\label{qmodes}
\left(
\begin{array}{c}
q_r \\
\bar q^{\dagger}_{r}\end{array}\right)=
\Biggl\{ -{1 \over 2} \left(
\begin{array}{cc}
 M_{rs} & {\widetilde M}_{rs} \\
-{\widetilde M}_{rs}& -M_{rs}\end{array}\right)+{ {\sqrt 3} \over 2}
\left(
\begin{array}{cc}
 0 & iC_{rs} \\
-iC_{rs}& 0 \end{array}\right) \Biggr\} \left(
\begin{array}{c} q_r \\
\bar q^{\dagger}_{r}\end{array}\right),
\end{equation}

\noindent
where the matrix $C$ is defined by

\begin{equation}
\label{cmatrix}
C_{rs}=(-1)^{r+1/2} \delta_{rs}.
\end{equation}

\noindent
This matrix implements BPZ conjugation and satisfies the following conditions:

\begin{eqnarray}
& C^2=1, \,\,\, C^T={\overline C}=C,\\
&CMC =M,\,\,\,\ C{\widetilde M}C =-{\widetilde M},\\
&CIC=-I,
\end{eqnarray}

\noindent
which guarantee the consistency of (\ref{qmodes}).

We now write the three string vertex as:

\begin{equation}
\label{threev}
|V_3 \rangle = \exp \Bigl[ {1 \over 2} q_3^{\dagger}\cdot I \cdot q_3
+ q^{\dagger} \cdot U \cdot \bar q^{\dagger}\Bigr] |0\rangle_{(123)},
\end{equation}

\noindent
where $I$ is the matrix (\ref{idenm}). This is of course a consequence of
(\ref{qthreecond}). Since $q|V_3 \rangle=U \bar q^{\dagger} |V_3 \rangle$,
by using (\ref{qmodes}) we obtain an explicit expression for $U$ in terms of 
$M$, $\widetilde M$ and $C$:

\begin{equation}
\label{Ue}
U = -{1 \over 2 +M} \cdot \bigl({\widetilde M} -i {\sqrt 3} C \bigr).
\end{equation}

\noindent
Using the above properties of $M$, $\widetilde M$ and $C$, it is easy to
show that $U$ satisfies,

\begin{eqnarray}
\label{propsu}
{\overline U} =-U^T =-C U C,\,\,\,\, IU={\overline U}I, \,\,\,\,\,
I{\overline U}=UI.
\end{eqnarray}

\noindent
The following formulae will also be useful:

\begin{eqnarray}
\label{formus}
I^2 &=& {M+1 \over M-1},\nonumber\\
U^2= {\overline U}^2 &=&
 {M-2 \over M+2},\nonumber\\
U + {\overline U}&=&-{2 {\widetilde M}  \over M+2}, \nonumber\\
U - {\overline U}&=&{2 {\sqrt 3} i C  \over M+2}.
\end{eqnarray}

\noindent
Let us now find the structure of the Neumann coefficients for the
three string vertex. These coefficients are defined through:

\begin{equation}
|V_3 \rangle = \exp \Bigl[ {1 \over 2} \eta_{\mu\nu} \sum 
\psi^{(a)\mu}_{-r} K^{ab}_{rs} \psi^{(b)\nu}_{-s} \Bigr] 
|0\rangle_{(123)},
\end{equation}

\noindent
and satisfy the condition

\begin{equation}
\label{antisym}
K_{rs}^{ab}=-K^{ba}_{sr}.
\end{equation}

\noindent
Using the above results, one finds that

\begin{equation}
\label{neustr}
K^{ab}={1 \over 3} \bigl( I + \omega^{b-a} U + \omega^{a-b} {\overline U}
\bigr),
\end{equation}

\noindent
which has the same structure as the Neumann coefficients in the bosonic 
sector. We also have the cyclicity property, $K^{a+1,
b+1}=K^{ab}$. We shall frequently use the matrices $K^{11}$, $K^{12}$ and
$K^{21}$, which are given by

\begin{eqnarray}
K^{11}&=&{1\over 3} (I + U + {\overline U}),\nonumber\\
K^{12}&=&{1\over 6}I -{1\over 6}(U + {\overline U}) +{i {\sqrt {3}} \over 6} 
(U-{\overline U}),\nonumber\\
K^{21}&=&{1\over 6}I -{1\over 6}(U + {\overline U}) -{i {\sqrt {3}} \over 6} 
(U-{\overline U}).
\end{eqnarray}

\noindent
Again, one can use conformal mapping techniques to write explicit
expressions for the Neumann coefficients \cite{GJ3}. The result is the
following. Define the coefficients $g_n$ through the expansion

\begin{equation}
\label{gseries}
\biggl( {1+ x \over 1-x}\biggr)^{1/6}= \sum_{n=0}^{\infty}
g_n x^n.
\end{equation}

\noindent
Next, define the following quantities:

\begin{eqnarray}
\label{Ms}
M^+_{nm} &=&-[(-1)^n -(-1)^m] [(n+1)g_{n+1} (m+1)g_{m+1}
-ng_n m g_m],\nonumber\\
M^-_{nm} &=&-[(-1)^n -(-1)^m] [ng_{n} (m+1)g_{m+1}
-(n+1)g_{n+1} m g_m],\nonumber\\
{\overline M}^+_{nm} &=&[(-1)^n +(-1)^m] [(n+1)g_{n+1} (m+1)g_{m+1}
-ng_n m g_m],\nonumber\\
{\overline M}^-_{nm} &=&[(-1)^n +(-1)^m] [ng_{n} (m+1)g_{m+1}
-(n+1)g_{n+1} m g_m].
\end{eqnarray}

\noindent
The Neumann coefficients are then given by,

\begin{eqnarray}
\label{Ks}
K^{aa}_{rs}&=&{1 \over 3} I_{rs} + i^{r+s} \biggl[ {M^+_{r-1/2,s-1/2}
\over r+s} + {M^-_{r-1/2,s-1/2}
\over r-s} \biggr],\nonumber\\
K^{aa+1}_{rs}&=&{1 \over 2} I_{rs}-{1 \over 2} K^{aa}_{rs} -{1\over 2}
{\sqrt 3}  i^{r+s-1} \biggl[ {{\overline M}^+_{r-1/2,s-1/2}
\over r+s} + {{\overline M}^-_{r-1/2,s-1/2}
\over r-s}\biggr],\nonumber\\
K^{aa-1}_{rs}&=&{1 \over 2} I_{rs}-{1 \over 2} K^{aa}_{rs} +{1\over 2}
{\sqrt 3}  i^{r+s-1} \biggl[ {{\overline M}^+_{r-1/2,s-1/2}
\over r+s} + {{\overline M}^-_{r-1/2,s-1/2}
\over r-s}\biggr].
\end{eqnarray}

\noindent
Using (\ref{neustr}), (\ref{Ks}), one can find explicit expressions for 
the matrices $U+\overline U$ and $U-\overline U$:

\begin{eqnarray}
\label{uus}
(U+ \overline U)_{rs}&=&3 i^{r+s}\biggl[ {M^+_{r-1/2,s-1/2}
\over r+s} + {M^-_{r-1/2,s-1/2}
\over r-s} \biggr],\nonumber\\
 (U-\overline U)_{rs} &=& 3 i^{r+s}\biggl[ {\overline M^+_{r-1/2,s-1/2}
\over r+s} + {\overline M^-_{r-1/2,s-1/2}
\over r-s} \biggr].
\end{eqnarray}

\noindent
Notice that the matrix $U-\overline U$ has nonzero diagonal terms. Using
the results of \cite{GJ3}, one finds,

\begin{equation}
\label{diaguu}
 (U-\overline U)_{rr}=6 i \Biggl[ {(n+1)^2 g_{n+1}^2 -n^2 g_n^2
\over 2n+1 } + {1 \over 3} \sum_{l=0}^n (-1)^l g_{n-l}^2\Biggr].
\end{equation}

\noindent
In the Appendix we show that these explicit expressions indeed agree 
with (\ref{Ue}).

For the calculations in the next section, it will be useful to define the 
following matrices

\begin{equation}
M^{ab}=CK^{ab}.
\end{equation}

\noindent
Using (\ref{neustr}) and the relations (\ref{propsu}), it is easy to see that
these matrices satisfy the following properties

\begin{equation}
\label{mcomm}
[M^{ab}, M^{a'b'}]=0, \,\,\,\,\,\,\,\,\, [CI, M^{ab}]=0.
\end{equation}

\noindent
These properties are of course similar to the properties of the matrices
$M^{ab}$ in the bosonic case \cite{KP,RSZclassical}.

\section{The Supersliver}

As we discussed in section 2, a factorized string field satisfies the
equations of motion of vacuum superstring field theory, with a pure ghost 
BRST operator, ${\cal Q}$, if the ghost part satisfies (\ref{ghosteq}) 
and the matter part is idempotent. We shall now consider idempotent 
matter states with the factorized form

\begin{equation}
\label{idemfact}
|\Psi\rangle =|\Psi_b\rangle \otimes |\Psi_f\rangle,
\end{equation}

\noindent
where $ |\Psi_{b,f}\rangle$ denote states which are obtained from
the vacuum by acting with bosonic and fermionic oscillators, respectively,
and which are idempotent with respect to the star product in their
respective matter sectors. In this section we will look for idempotent 
states in the fermionic sector. First, we provide an algebraic construction, 
in the spirit of \cite{KP}. Then we compare the solution to the geometric 
construction of the sliver given in \cite{RZ}.

\subsection{Algebraic Construction}

Our purpose here is to find a state in the fermionic part of the
matter sector that star squares to itself. Our ansatz is a squeezed state
of the form

\begin{equation}
\label{psif}
|\Psi_F \rangle = {\cal N}_F \exp \Bigl[ -{1 \over 2} \eta_{\mu \nu}
\sum_{r,s \ge {1\over 2}} \psi^{\mu}_{-r} F_{rs} \psi^{\nu}_{-s} \Bigr]
|0 \rangle,
\end{equation}

\noindent
where $F_{rs}$ is an antisymmetric matrix. Recall that the star product 
of two states, $|\Psi\rangle$, $|\Phi \rangle$, defined as

\begin{equation}
\label{star}
|\Psi \star \Phi \rangle_{(3)}= {}_{(1)}\langle \Psi| {}_{(2)} \langle
 \Phi | | V_3 \rangle_{(123)},
\end{equation}

\noindent
involves the BPZ conjugate of the string field states. To obtain the BPZ 
conjugate of $|\Psi_F \rangle$, one has to take into account that

\begin{equation}
{\rm bpz}(\psi^{\mu}_r)=(-1)^{r+1/2}\psi^{\mu}_{-r}.
\end{equation}

\noindent
Therefore, the matrix that implements BPZ conjugation is $C$. It will be
useful in the following to define:

\begin{equation}
H=CF.
\end{equation}

\noindent
In order to evaluate the star product, one still needs the following formula. 
Let $b_i$, $b_i^{\dagger}$ be fermionic oscillators with anticommutation 
relations $\{b_i, b_j^{\dagger} \}=\delta_{ij}$, let $\lambda_i$, 
$\mu_i$ be a set of Grassmann variables, and let $S_{ij}$, $T_{ij}$ be 
antisymmetric matrices. One then has

\begin{eqnarray}
\label{bosc}
& &\langle 0| \exp \Bigl( \lambda^T \cdot b +
{1 \over 2} b \cdot S \cdot b \Bigr)
 \exp \Bigl( \mu^T \cdot b^{\dagger} + {1 \over 2}
b^{\dagger} \cdot T
\cdot b^{\dagger} \Bigr)|0 \rangle = \nonumber\\
&=&\Bigl[{\rm det} (1 + ST)\Bigr]^{1/2} \exp\Bigl[ \mu^T (1 +ST)^{-1}
 \lambda + {1 \over 2} \lambda^T T (1+ST)^{-1} \lambda +
{1 \over 2} \mu^T (1+ST)^{-1}S\mu\Bigr].
\end{eqnarray}

\noindent
Similar expressions for bosonic oscillators and for the ghost $bc$ system 
were presented in \cite{KP,RSZclassical}. Using this 
formula, one obtains the following
expression:

\begin{eqnarray}
|\Psi_F \star \Psi_F \rangle_{(3)} &=&{\cal N}_F^2
\Bigl[{\rm det} (1 + \Phi {\cal K})\Bigr]^{5} \cdot \nonumber\\
& \cdot& \exp\Bigl[ {1 \over 2} \eta_{\mu \nu} \bigl\{
\chi^{\mu T} (1+ \Phi {\cal K})^{-1}\Phi \chi^{\nu} + {1 \over 2}
\psi^{3\mu}_{-r}K^{33}_{rs}\psi^{3\nu}_{-s}\bigr\} \Bigr]
|0\rangle_{(3)},
\end{eqnarray}

\noindent
where

\begin{equation}
\label{matpod}
\Phi=\left( \begin{array}{cc} -HC & 0 \\
0 &- HC \end{array} \right),\,\,\,\,\,\,
{\cal K}=\left( \begin{array}{cc} K^{11} & K^{12} \\
K^{21} & K^{22} \end{array} \right),\,\,\,\,\,\,
\chi^{\mu}=\left( \begin{array}{c} K^{13}\psi^{3\mu\dagger} \\
K^{23}\psi^{3\mu\dagger}\end{array}\right).
\end{equation}

\noindent
Using (\ref{antisym}) and the cyclicity property, one further obtains the 
following equation for $H$,

\begin{equation}
H=-M^{11} - \left( \begin{array}{cc} M^{12} & M^{21} \end{array} \right)
\left( \begin{array}{cc} 1-HM^{11} & -HM^{12} \\
-HM^{21} & 1 - HM^{22} \end{array}\right)^{-1}
\left( \begin{array}{c} HM^{21}\\
HM^{12}\end{array}\right),
\end{equation}

\noindent
and the following value for the normalization constant,

\begin{equation}
\label{norm}
{\cal N}_F= \bigl[ {\rm det}(1 + \Phi {\cal K})\bigr]^{-5}.
\end{equation}

\noindent
Since the matrices $M^{ab}$ commute, one can assume that $[H,M^{ab}]=0$ and
proceed as if we were dealing with commuting variables. After some simple
algebra, one finds the following cubic equation for $H$:

\begin{equation}
\label{cubic}
A_3 H^3 + A_2 H^2 + A_1 H + A_0 =0,
\end{equation}

\noindent
where

\begin{eqnarray}
\label{coeffs}
A_3 &=&M^{12}M^{21} -(M^{11})^2 ,\nonumber\\
A_2&=&3 M^{11}M^{12}M^{21}-(M^{11})^3- (M^{12})^3 -(M^{21})^3,\nonumber\\
A_1 &=& -1 -2 A_3,\nonumber\\
A_0&=&-M^{11}.
\end{eqnarray}

\noindent
In the bosonic case analyzed in \cite{KP} and \cite{RSZclassical},
the coefficients of the cubic equation for the bosonic piece of the sliver
could be simplified by using relations among the matrices of Neumann
coefficients. Here, it is convenient to express (\ref{coeffs}) in terms of
the matrices $U$, ${\overline U}$ and $I$, which in turn can be expressed
in terms of $C$, $M$ and ${\widetilde M}$. After some simple algebra, one
finds the following:

\begin{eqnarray}
A_3&=&{1 \over 3} (U + {\overline U})I + {1 \over 6} (U^2 + {\overline
U}^2) =\nonumber \\
&=& {M \over M+2},\nonumber\\
A_2&=&{1 \over 2}C(U^2 + {\overline U}^2)I +
{2\over 3}C(I + U + {\overline U}) =\nonumber\\
&=& -C {\widetilde M} { 3M -2 \over (M-1)(M+2)},\nonumber\\
A_1&=&-{3M +2 \over M+2},\nonumber \\
A_0&=&C{\widetilde M} {M \over (M-1) (M+2)}.
\end{eqnarray}

\noindent
Since $|I \star I \rangle= |I \rangle$, an important check of the above 
is wether $H=-CI$ is a solution of (\ref{cubic}). In fact,
one can further write (\ref{cubic}) as

\begin{equation}
\label{quadratic}
(H+CI)(M H^2 -2 C {\widetilde M}H - M) =0.
\end{equation}

\noindent
In order to solve the quadratic equation, one has to be careful when 
extracting the square root. Since $F$ must be antisymmetric, and remembering 
that $F=CH$, one finds the two solutions,

\begin{equation}
\label{sols}
F^{\pm}={ {\widetilde M} \over M} \biggl( 1 \pm {1 \over {\sqrt {1 -M^2}}}
\biggr).
\end{equation}

\noindent
Notice that $CF^{\pm}C=-F^{\pm}$. It is also easy to check that $H$ commutes 
with $M^{ab}$, as assumed in our initial ansatz. Using the above result for 
$F^{\pm}$, one can compute

\begin{equation}
(1+ \Phi {\cal K})^{-1}= -{1 \over 4}(M-1)(M+2)\Bigl( 1 \mp {M+1 \over
 {\sqrt {1 -M^2}}}\Bigr),
\end{equation}

\noindent
which determines the normalization constant ${\cal N}_{F^{\pm}}$ through
(\ref{norm}). Using again (\ref{bosc}), one can further compute the norm
of $|\Psi^{F_{\pm}}\rangle$ and find, for both signs,

\begin{equation}
\label{norma}
\langle \Psi^{F_{\pm}} | \Psi^{F_{\pm}} \rangle =
\Bigl[ {\rm det} ((1-M)(1+M/2)^2) \Bigr]^5.
\end{equation}

\noindent
Finally, notice that in order for the identity to star square to itself, 
one needs

\begin{equation}
\label{normid}
{\cal N}_I= \Bigl[ {\rm det} ((1-M)(1+M/2))\Bigr]^5,
\end{equation}

\noindent
and its BPZ norm turns out to be

\begin{equation}
\langle I | I \rangle = \Bigl[ {\rm det} (2(1-M)(1+M/2)^2)\Bigr]^5.
\end{equation}

\subsection{Numerical Results and Comparison to the Geometric Sliver}

The above results involve infinite--dimensional matrices. They can 
however be analyzed numerically by restricting the matrix rank to $L 
< \infty$ and then using suitable numerics in order to study the limit 
$L \rightarrow \infty$, as in \cite{RSZclassical}. The first thing to notice is 
that the determinant of $M$ converges to zero very rapidly. As a consequence, 
the solution $F^{+}$, which behaves like $2{\widetilde M} M^{-1}$, has diverging 
eigenvalues. The other solution, which behaves like ${\widetilde M}M/2$, 
has a better behavior. This is the solution that we will discuss in the
rest of the paper, and we shall henceforth simply denote it by $F=F^-$. 

It turns out that $F$ is the matrix that appears in the fermionic part of the
geometric sliver constructed by Rastelli and Zwiebach in \cite{RZ}. Since
the sliver can be defined purely in geometric terms, one can construct a
supersliver in the CFT given by the NS sector of the superstring. Recall
that the (super)sliver is defined by

\begin{equation}\label{ss}
\langle \Xi | =\langle 0|U_f,
\end{equation}

\noindent
where $U_f$ is the operator associated to the conformal transformation
given by

\begin{equation}
f(z) =\arctan (z).
\end{equation}

\noindent
The structure of the operator $U_f$ was found in \cite{RZ}. It is given by:

\begin{equation}
U_f ={\rm e}^{\sum_{n=1}^{\infty} a_n L_{-2n}},\end{equation}

\noindent
where the coefficients $a_n$ can be computed explicitly. The Virasoro
operators split as $L=L_b+L_f+L_g$, where $b$, $f$, $g$ refer respectively
to the bosonic matter, fermionic matter and ghost/superghost sectors. As a
consequence, the supersliver will factorize as:

\begin{equation}
|\Xi\rangle =  |\Xi_b\rangle \otimes |\Xi_f\rangle \otimes |\Xi_g\rangle.
\end{equation}

\noindent
The bosonic matter part is the one constructed algebraically in \cite{KP}.
In the following, we will present evidence that the fermionic matter
part is the idempotent state constructed above and corresponding to $F$,
{\it i.e.},

\begin{equation}
|\Xi_f\rangle = |\Psi_{F} \rangle.
\end{equation}

\noindent
The first step is, as in \cite{RSZclassical}, to write $|\Xi_f\rangle$ 
as a squeezed state:

\begin{equation}
|\Xi_f\rangle={\cal N} \exp\Bigl[ -{1 \over 2}\eta_{\mu\nu} \sum_{r,s}
\psi^{\mu}_{-r} {\widehat F}_{rs} \psi^{\nu}_{-s}\Bigr] |0\rangle.
\end{equation}

\noindent
Using the CFT techniques of \cite{LPP, LPPtwo, LC}, one finds for the 
matrix $\widehat{F}$:

\begin{equation}
\label{doubleres}
\widehat F_{rs}=-\oint_0 {dw \over 2\pi i}
\oint_0 {dz \over 2\pi i}
{z^{-r-1/2} w^{-s-1/2} \over (1+z^2)^{1/2} (1 + w^2 )^{1/2}
(\tan^{-1}(z)-
\tan^{-1}(w))}.
\end{equation}

\noindent
One can see that $\widehat F_{rs}=0$ if $r+s={\rm odd}$, {\it i.e.},
$C\widehat FC=-\widehat F$, as follows from the algebraic 
description. Evaluating the residues, one finds for the first nonzero 
entries:

$$
\widehat F_{1/2,3/2}=-{1 \over 6} \simeq -.1666, \,\,\,\,\,\,\,\,\, 
\widehat F_{1/2, 7/2}={43\over 60} \simeq 0.1194, \,\,\,\,\,\,\,\,\, 
\widehat F_{1/2,11/2}=-{1459\over 15120} \simeq -0.0964,
$$

\begin{equation}
\label{residua}
\widehat F_{3/2,5/2}=-{1 \over 40} \simeq -.0250, \,\,\,\,\,\,\,\,\,
\widehat F_{3/2,9/2}={71\over 15120} \simeq .0046,  \,\,\,\,\,\,\,\,\,
\widehat F_{5/2, 7/2}=-{ 239 \over 7560} \simeq -.0316.
\end{equation}

\noindent
On the other hand, we can evaluate numerically the first few coefficients
$F_{rs}$. Since $M, \widetilde M$ do not commute at finite rank,
we can approximate the matrix $F$ in two ways: multiplying ${\widetilde M}$
on the right, or on the left. The results are shown, respectively, in the
following tables:

 \[
\begin{array}{|c|r|r|r|r|r|r|}  \hline
L
& F_{1/2,3/2}           & F_{1/2,7/2}                      &
F_{1/2,11/2}                        & F_{3/2,5/2}                      &
F_{3/2,9/2} & F_{5/2,7/2}      \\
\hline
20& $-0.1929$ & $0.1427$ &  $-0.1186$ &  $0.0033$ & $-0.0178$&$-0.0448$   \\
\hline
100 & $-0.1876$ & $0.1347$ & $-0.1102$ & $-0.0058$ & $-0.0099$ &$-0.0398$\\
\hline
150 & $-0.1847$ & $0.1335$ & $-0.1091$ & $-0.0074$ & $-0.0087$&$-0.0391$ \\
\hline
\infty & $-0.1676$ & $0.1235$ & $-0.1098$ & $-0.0268$ & $0.0036$ & $-0.0347$
\\
\hline
\end{array}
\]

\bigskip

\[
\begin{array}{|c|r|r|r|r|r|r|}  \hline
L
& F_{1/2,3/2}           & F_{1/2,7/2}                      &
F_{1/2,11/2}                        & F_{3/2,5/2}                      &
F_{3/2,9/2} & F_{5/2,7/2}      \\
\hline
20& $-0.1140$ & $0.0752$ &  $-0.0570$ &  $-0.0397$ & $0.0163$&$-0.0076$  \\
\hline
100 & $-0.1299$ & $0.0886$ & $-0.0683$ & $-0.0346$ & $0.0122$ & $-0.0155$  \\
\hline
150 & $-0.1328$ & $0.0910$ & $-0.0710$ & $-0.0338$ & $0.0115$ & $-0.0168 $   \\
\hline
\infty & $-0.1726$ & $0.1251$ & $-0.1020$ & $-0.0250$ & $0.0045$& $-0.0335$  \\
\hline
\end{array}
\]

\noindent
The last entry shows an extrapolation to $L=\infty$ by fitting fifteen
points $L=10,20, \dots, 150$ to $a_0 + a_1/(\log L) + a_2 (\log L)^2$.
We see that there is good agreement with the exact result
(\ref{residua}), and this provides good numerical evidence that the matrix 
$F$ is indeed given by the double residue (\ref{doubleres}).

It is also interesting to consider the behavior of the BPZ norms of 
the fermionic identity and the fermionic part of the sliver. The fermionic 
identity turns out not to be normalizable: the determinant in (\ref{normid}) 
grows very quickly as we increase the rank. On the other hand, the norm of 
$|\Xi_f\rangle$, given in (\ref{norma}), behaves like the norm of 
$|\Xi_b\rangle$ analyzed in \cite{RSZclassical}: an extrapolation
to infinite rank, by fitting one hundred points $L=10, 20, \dots, 1000$,
gives $\langle \Xi_f|\Xi_f\rangle^{1/5}=-0.0075$. This seems to
indicate that the norm of the fermionic part of the supersliver is zero.

\subsection{Conservation Laws}

In this subsection we wish to derive conservation laws satisfied by the 
supersliver, involving the superconformal generators, $G_{r}$, and following 
\cite{RZ, RSZclassical}. We shall be schematic, as the procedure is by now 
well known. Observe that due to its purely geometrical construction 
(\ref{ss}) the supersliver will clearly satisfy all the Virasoro conservation 
laws outlined in the Appendix of \cite{RSZclassical}, involving the $L_{n}$ 
generators of the conformal algebra which now will have a fermionic 
matter piece as well as a bosonic and ghost pieces. Let us then 
outline how can one derive the conservation laws associated to the 
rest of the superconformal algebra, \textit{i.e.}, the ones 
depending on the $G_{r}$ generators.

The sliver surface state is defined in the upper half plane by the 
conformal map,

\begin{equation}
f_{\mathrm{H}} (z) = \arctan \left( z \right),
\end{equation}

\noindent
while in the unit disk (coordinates that we will use in the 
following), it is given by

\begin{equation}
f_{\mathrm{U}} (z) = \frac{1+i\arctan \left( z \right)}{1-i\arctan 
\left( z \right)}. 
\end{equation}

\noindent
The usual contour deformation argument yields the expected 
conservation law,

\begin{equation}
\langle \Xi | \oint dz\ \varphi(z)\ G(z) = 0,
\end{equation}

\noindent
where $G(z)$ is the super stress tensor, $G(z) = \sum G_{r} / 
z^{r+\frac{3}{2}}$, and the conformal densities $\varphi(z)$ now have 
weight $-1/2$. Precisely because of this non--integer weight, one has 
to be careful when taking the conformal transformation,

\begin{equation}
\varphi(z) = \tilde{\varphi} \left( f(z) \right) \left( \frac{d f 
\left( z \right)}{d z} \right)^{-\frac{1}{2}},
\end{equation}

\noindent
so that we shall adopt the standard conventions \cite{thesis}.

With the choice of conformal density,

\begin{equation}
\varphi(z) = -\frac{4}{3} \sqrt{\frac{2}{3}} \left( 1 - i \right)
\left( 1 + \frac{1}{z-1} \right),
\end{equation}

\noindent
one obtains the following conservation law,

\begin{equation}
\langle \Xi | \left( G_{-3/2} + \frac{11}{6}\ G_{1/2} + \frac{43}{360}\ 
G_{5/2} - \frac{1039}{15120}\ G_{9/2} + \cdots \right) = 0.
\end{equation}

\noindent
If instead one chooses the conformal density,

\begin{equation}
\varphi(z) = -\frac{4}{3} \sqrt{\frac{2}{3}} \left( 1 - i \right) 
\left( \frac{1}{2} + i\ \frac{\sqrt{3}}{2} - \frac{1}{z-e^{\frac{2\pi 
i}{3}}} \right),
\end{equation}

\noindent
one obtains the conservation law,

\begin{equation}
\langle \Xi | \left( G_{-1/2} + \frac{1}{\sqrt{3}}\ G_{1/2} + 
\frac{11}{6}\ G_{3/2} + \frac{3}{2\sqrt{3}}\ G_{5/2} + \frac{7}{72}\ 
G_{7/2} + \cdots \right) = 0.
\end{equation}

\noindent
Other conservation laws can be obtained in similar fashions.

\section{Fermionic Star Algebra Spectroscopy}

In this section we follow the methods of \cite{RSZspectro}, in order to 
find the eigenvalue spectrum of the various infinite--dimensional matrices 
involved in the star algebra for the matter fermionic sector, as 
well as the corresponding eigenvectors. We first find by inspection 
an eigenvector of $M$ and $\widetilde M$, and we then adapt the methods of 
\cite{RSZspectro} to find the rest of the spectrum. The star algebra
spectroscopy has also been studied in \cite{HM, MT}.

\subsection{An Eigenvector of $M$ and $\widetilde M$}

In this subsection we want to show that the matrices $M$ and $\widetilde 
M$ have a common eigenvector with eigenvalues $-1$ and $0$, respectively. 
First define

\begin{equation}
\label{vectorv}
\nu_{n-1/2}=\begin{cases} \left( \begin{array}{c} -1/2 \\
k \end{array}\right),& n=2k+1, \\
0, & n=2k. \end{cases}
\end{equation} 

\noindent
Using (\ref{sumbin}), and setting $r=n+1/2$, one easily finds

\begin{eqnarray}
\sum_{s}M_{rs}\nu_s & =& -{2 \over \pi} \sum_{m=0}^{\infty} 
{(-1)^{n-m} \over 2m+2n+1}\left( \begin{array}{c} -1/2 \\
m \end{array}\right)=-\left( \begin{array}{c} -1/2 \\
n \end{array}\right)=-\nu_r,\nonumber\\
\sum_{s}{\widetilde M}_{rs}\nu_s & =& {2 \over \pi} \sum_{m=0}^{\infty} 
{(-1)^{n+m} \over 2m-2n-1}\left( \begin{array}{c} -1/2 \\
m \end{array}\right)={(-1)^n \over \pi} 
{\Gamma({1 \over 2}) \Gamma (-n-{1\over 2}) \over  
\Gamma (-n)}=0.
\end{eqnarray}

\noindent
Therefore, $\nu_{r}$ is a common eigenvector to $M$ and $\widetilde M$ 
with eigenvalues $-1$ and $0$, respectively. This vector can be understood 
geometrically as follows. Notice that its components are the negative modes 
in the Fourier expansion of the function 

\begin{equation}
f(\sigma)= { {\rm e}^{-i {\sigma \over 2}} \over 
{\sqrt {1 + {\rm e}^{-2i \sigma}}}}.
\end{equation}

\noindent
This function is antiperiodic in $[-\pi, \pi]$ and satisfies the overlap 
equation $f(\sigma)=if(\pi -\sigma)$. Therefore, its modes satisfy the 
equation (\ref{mmodes}). Since the positive modes are set to zero, it 
follows from (\ref{mmodes}) that the coefficients of the Fourier expansion 
give an eigenvector of $M$ and $\widetilde M$ with the required eigenvalues. 
Finally, notice that the vector $\nu$ is BPZ odd, since $C \nu=-\nu$. A 
related discussion of the geometric meaning of the eigenvectors in the 
bosonic case can be found in \cite{MT}.

\subsection{Diagonalizing $K_1$}

To find the rest of the spectrum, we generalize the considerations of 
\cite{RSZspectro} to the fermionic sector. The derivation of the 
star algebra,

\begin{equation}
K_1=L_1 + L_{-1},
\end{equation}

\noindent
has a fermionic part which is a sum of bilinears in the modes $\psi_{\pm
r}$. This allows for a definition of an infinite--dimensional matrix as 
follows. Let $\{ v_r\}_{r \ge 1/2}$ be an infinite--dimensional vector. 
Define then the matrix ${\rm K}_1$ through

\begin{equation}
[K_1, v\cdot \psi]=({\rm K}_1 v) \cdot \psi,
\end{equation}

\noindent
where $v\cdot \psi =\sum_{r=1/2}^{\infty} v_r \psi_r$. In what follows, 
it will be quite useful to label the positive half--integer indices with 
integer numbers by setting $r=n-1/2$, $n=1,2, \cdots$. Using the explicit
expression for the Virasoro generators, we then find:

\begin{equation}
({\rm K}_1)_{nm}=-(n-1) \delta_{n-1,m} -n \delta_{n+1,m}.
\end{equation}

\noindent
This is a symmetric matrix that anticommutes with $C$, $\{{\rm K}_1,C
\}=0$. To find its spectrum, one associates to every vector $w$ a function
$f_w(z)$ as follows

\begin{equation}
f_w (z)=\sum_{n=1}^{\infty} w_n z^n.
\end{equation}

\noindent
The infinite--dimensional matrix ${\rm K}_1$ is then represented in the space 
of functions by the differential operator

\begin{equation}
{\cal K}_1= -(1+z^2) {d \over dz} +{1 \over z},
\end{equation}

\noindent
and the problem of finding eigenvectors of ${\rm K}_1$ now becomes the problem
of finding eigenfunctions for this differential operator. The solution is
immediate: for any $-\infty < \kappa < \infty$ there is an eigenfunction of
${\cal K}_1$ given by

\begin{equation}
\label{wk}
f_{w^{(\kappa)}}(z)= {z \over {\sqrt {1+z^2}}} \exp \bigl( - \kappa\
\arctan(z) \bigr),
\end{equation}

\noindent
with eigenvalue $\kappa$. The normalization of this function
has been chosen so that $w^{(\kappa)}_1=1$. One then concludes that ${\rm
K}_1$ has a non--degenerate, continuous spectrum, similar to the bosonic
case studied in \cite{RSZspectro}. Also notice that 

\begin{equation}
f_{Cw}(z)=-f_w (-z),
\end{equation}

\noindent
so that the BPZ matrix acts as

\begin{equation}
C w^{(\kappa)}=-w^{(-\kappa)}.
\end{equation}

\subsection{Diagonalizing $M$ and ${\widetilde M}$}

We can now use this information in order to find the spectrum of $M$ 
and $\widetilde M$. First, observe the following properties,

\begin{eqnarray}
\label{derivatio}
\left[{\rm K}_1, CI \right]&=&0,\nonumber\\
\left[{\rm K}_1, M^{ab}\right]&=&0.
\end{eqnarray}

\noindent
The first equation follows from the fact that $K_1$ kills the identity, and 
the second one from the fact that $K_1$ is a derivation of the star
algebra, and then $(K^{(1)}_1+ K^{(2)}_1+ K^{(3)}_1)|V_3 \rangle =0$ 
\cite{witssft}. To derive (\ref{derivatio}), we have also used the fact that 
${\rm K}_1$ anticommutes with $C$. Making use of (\ref{neustr}), it 
follows that

\begin{equation}
[{\rm K}_1, M]=[{\rm K}_1, C {\widetilde M}]=0.
\end{equation}

\noindent
Therefore, and since the spectrum of ${\rm K}_1$ is nondegenerate, an
eigenvector of ${\rm K}_1$ has to be an eigenvector of $M$ and 
$C \widetilde M$ as well. Notice that this makes sense since $M$ and 
$C \widetilde M$ are symmetric, real matrices, and so they have real 
eigenvalues. 

We have then shown that the eigenvectors $w^{(\kappa)}$ given
implicitly in (\ref{wk}) are also eigenvectors of $M$ and $C \widetilde
M$. Now, we have to find out which are the corresponding eigenvalues. This 
can be done with a trick from section 5.2 of \cite{RSZspectro}. The
eigenvalue equations are

\begin{eqnarray}
M_{n-1/2,\, m-1/2} w^{(\kappa)}_{m}& = & m(\kappa) w^{(\kappa)}_{n},
\nonumber\\
( C{\widetilde M})_{n-1/2,\, m-1/2} w^{(\kappa)}_{m}& = & \widetilde
m(\kappa) w^{(\kappa)}_{n}.
\end{eqnarray}

\noindent
Since we chose the normalization $w^{(\kappa)}_{1}=1$, one can consider the 
above equations with $n=1$ and obtain for the eigenvalues:

\begin{eqnarray}
m(\kappa)&=& {2 \over \pi} \sum_{q=1}^{\infty} {(-1)^q \over 2q-1} 
w_{2q-1}^{(\kappa)},\nonumber\\  
\widetilde m(\kappa)& =&- {2 \over \pi} 
\sum_{q=1}^{\infty} {(-1)^q \over 2q-1} 
w_{2q}^{(\kappa)}.
\end{eqnarray}

\noindent
Define now the functions 

\begin{eqnarray}
\mu(z)&=& \sum_{q=1}^{\infty} {(-1)^q \over 2q-1} 
w_{2q-1}^{(\kappa)}z^{2q-1},\nonumber\\  
\widetilde \mu(z)& =&\sum_{q=1}^{\infty} {(-1)^q \over 2q-1} 
w_{2q}^{(\kappa)} z^{2q-1},
\end{eqnarray}

\noindent
which can be found to satisfy

\begin{eqnarray}
{ d\mu (z)\over dz}&=& {i \over 2z}
(f_{w^{(\kappa)}}(iz)-f_{w^{(\kappa)}}(-iz)), \nonumber\\
\,\,\,\,\,\,\,\, { d\tilde \mu (z)\over dz}&=&{i \over 2z^2}
(f_{w^{(\kappa)}}(iz)+f_{w^{(\kappa)}}(-iz)).
\end{eqnarray}

\noindent
Using the explicit expression for $f_{w^{(\kappa)}}(z)$, the fact 
that $\mu(0)=\tilde \mu (0)=0$, and integrating, one finally finds

\begin{eqnarray}
\mu (1)&=&-{\pi \over 2} {\rm sech} (\kappa \pi/2),\nonumber\\
\widetilde \mu (1)&=&{\pi \over 2} \tanh (\kappa \pi/2).
\end{eqnarray}

\noindent
This determines the eigenvalues of $M$ and $C\widetilde M$ for the 
eigenvectors $w^{(\kappa)}$:

\begin{eqnarray}
\label{finaleigen}
m(\kappa)&=& -{\rm sech} (\kappa \pi/2),\nonumber\\   
\widetilde m(\kappa)& =&-\tanh  (\kappa \pi/2).
\end{eqnarray}

\noindent
The spectrum of $M$ lies in the interval $[-1,0)$, while that of $C
\widetilde M$ lies in $(-1,1)$. The above results are of course compatible 
with the relation $M^2 -\widetilde M^2=1$. Notice finally that, for $\kappa=0$, 
we recover the results of the previous subsection, since

\begin{equation}
f_{w^{(0)}}(z) = {z \over {\sqrt {1 + z^2}}}= \sum_{n=0}^{\infty} 
\left( \begin{array}{c} -1/2 \\
n \end{array}\right) z^{2n+1},
\end{equation}

\noindent
so $w^{(0)}=\nu$ and from (\ref{finaleigen}) we read that the eigenvalues 
with respect to $M$ and ${\widetilde M}$ are in fact $-1$ and $0$, 
respectively, in agreement with the explicit computations of the previous 
subsection. 

We can now diagonalize the symmetric matrix $H=CF$ that defines the fermionic 
sliver. Since the derivation $K_1$ kills the supersliver \cite{RSZclassical}, 
one has that 

\begin{equation}
[{\rm K}_1, H]=0,
\end{equation}

\noindent
and by the same argument one has that $w^{(\kappa)}$ are eigenvectors of 
$H$. The corresponding eigenvalues will be denoted by $h(\kappa)$. In 
order to determine them first notice that, since $H$ anticommutes with $C$, 
one has

\begin{equation}
\label{oddh}
h(\kappa)=-h(-\kappa).
\end{equation}

\noindent
We can determine $h(\kappa)$ from the explicit expression given in 
(\ref{sols}). However, one has to be careful when doing this. The reason 
is that $\widetilde M/(1-M^2)^{1/2}$ gives an indeterminacy of the 
type $0/0$ when acting on $w^{(0)}$. If one naively substitutes the 
eigenvalues in (\ref{sols}), one seems to find that $h(0)\not=0$, which 
contradicts (\ref{oddh}). Of course the appropriate way to regularize 
this indeterminacy is by expanding $(1-M^2)^{-1/2}$ in powers of $M$, and 
if this is done then at every order in the expansion one indeed finds the 
right value of the eigenvalue, which is $h(0)=0$ (and can also be checked by 
computing $H w^{(0)}$ in level truncation). Related issues associated to the 
appearance of inverses of singular matrices have been considered in the bosonic 
case in \cite{HM}. Another subtlety (also present in the bosonic case analyzed 
in \cite{RSZspectro}) is that the quadratic equation determining $H$ gives 
two branches for the eigenvalues, and in fact there is a jump from one 
branch to the other at $\kappa=0$. Since the numerical analysis of the 
spectrum shows that the eigenvalues of $H$ are in the interval $[-1,1]$, 
one finally finds that the spectrum of $H$ is given by

\begin{equation}
\label{hvalue}
h(\kappa)=\begin{cases}
-{\kappa \over |\kappa|} {\rm e}^{-|\kappa| \pi/2},& \kappa \not= 0,  \\
 0 ,& \kappa =0,
\end{cases}
\end{equation}     

\noindent
in agreement with (\ref{oddh}). 

Using all these results, one can also diagonalize the rest of the matrices 
that we have encountered so far. For example, the eigenvalues of the real 
symmetric matrices $M^{11}, M^{12}, M^{21}$ are, respectively,

\begin{eqnarray}
m^{11}(\kappa)&=&-{ \sinh (\kappa \pi/2) \over 
(1+ \cosh (\kappa \pi/2))(1-2\cosh (\kappa \pi/2))},\nonumber\\
m^{12}(\kappa)&=&{ \cosh (\kappa \pi/2) \bigl( 1+\cosh (\kappa \pi/2)+ 
\sinh (\kappa \pi/2) \bigr)\over 
(1+ \cosh (\kappa \pi/2))(1-2\cosh (\kappa \pi/2))},\nonumber\\
m^{21}(\kappa)&=&-{ \cosh (\kappa \pi/2) \bigl( 1+\cosh (\kappa \pi/2)- 
\sinh (\kappa \pi/2) \bigr)\over 
(1+ \cosh (\kappa \pi/2))(1-2\cosh (\kappa \pi/2))}.
\end{eqnarray}

\noindent
Of course there is still the possibility that all of these matrices have 
other eigenvectors which are not eigenvectors of ${\rm K}_1$. We have 
not performed a systematic numerical search, but we are led believe that, 
just as in the bosonic case, the above results determine the complete 
spectrum of eigenvectors and eigenvalues of the various infinite--dimensional 
matrices involved in the fermionic matter sector.

It is also interesting to observe that, again just as in the bosonic case
\cite{MT, RSZspectro}, the eigenvectors that we have found are not
normalizable. This can be seen in detail as follows. Given two
infinite--dimensional vectors $v$ and $w$, their inner product is given by

\begin{equation}
\label{fnorm}
v\cdot w \equiv\sum_{n=1}^{\infty} v_n w_n= \int_0^{2\pi} {d \theta \over
2\pi} f_v^*({\rm e}^{i \theta}) f_w ({\rm e}^{i \theta}). 
\end{equation}

\noindent
The norm of a vector $v$ is defined as usual by $\| v\|^2 \equiv v\cdot v$. 
Using (\ref{wk}) and (\ref{fnorm}), one can find an explicit expression for
the norm of $w^{(\kappa)}$, 

\begin{equation}
\label{wnorm}
\| w^{(\kappa)} \|^2 = \cosh (\kappa \pi/2) \| \nu \|^2,
\end{equation}

\noindent
where

\begin{equation} 
\| \nu \|^2= 4
\int_0^{\pi/2} {d\theta \over 2\pi} {1\over {\sqrt {2 + 2 \cos (2
\theta)}}}.
\end{equation}

\noindent
This integral is logarithmically divergent, so the norm of $\nu$ (and
therefore of all $w^{(\kappa)}$) is infinite. Another way to see
this is to compute directly the sum:

\begin{equation}
\| \nu \|^2=\sum_{n=0}^{\infty}\left( \begin{array}{c} -1/2 \\
n \end{array}\right)^2.
\end{equation}

\noindent
By using zeta--function regularization, we find that this 
series diverges as

\begin{equation}
\lim_{\epsilon \to 0}\ \frac{2}{\pi}\ K 
\left( \mathrm{e}^{-\epsilon} \right),
\end{equation}
 
\noindent
where $K(x)$ is the elliptic $K$--function, which indeed diverges
logarithmically as $x\rightarrow 1$.

\section{Coherent States and Higher--Rank Projectors}

Once the fermionic sliver has been constructed, it is natural to consider
fermionic coherent states based on it, in analogy to the bosonic case
\cite{KP, RSZhalf}. In this section we shall construct coherent states 
and determine their star products. This will be useful in order to construct 
higher--rank projectors of the fermionic star algebra---idempotent states 
that should represent multiple $D$--brane configurations \cite{RSZhalf}.

\subsection{Coherent States on the Supersliver}

We define fermionic coherent states as follows. Let $\beta =\{\beta \}_r$, 
$r \ge 1/2$, be a Grassmannian vector. Then, the coherent state on the 
fermionic sliver associated to $\beta$, that we shall denote by 
$| \Xi_{\beta} \rangle$, is given by

\begin{equation}
\label{cohe}
| \Xi_{\beta} \rangle =\exp \bigl[ (-C \beta)^T \cdot
\psi^{\dagger}\bigr] |\Xi_f \rangle. 
\end{equation}

\noindent
This definition guarantees that the BPZ conjugate of (\ref{cohe}) has a 
simple expression, 

\begin{equation}
\langle \Xi_{\beta} | =\langle \Xi_f | \exp \bigl[  \beta^T \cdot
\psi \bigr].
\end{equation}

\noindent
The star product of two coherent states can be computed very easily by 
using (\ref{bosc}), and one finds

\begin{equation}
| \Xi_{\beta_1} \rangle \star | \Xi_{\beta_2} \rangle = 
\exp \biggl( \chi^T (1+ \Phi {\cal K})^{-1} \beta +{1 \over 2}\beta^T 
{\cal K}(1+ \Phi {\cal K})^{-1} \beta \biggr) |\Xi_f\rangle,
\end{equation}

\noindent
where $\Phi$, $\cal K$ and $\chi$ are given in (\ref{matpod}), and 
$\beta$ denotes here the vector

\begin{equation}
\beta =\left( \begin{array}{c} \beta_1  \\
\beta_2 \end{array}\right).  
\end{equation}

\noindent
An explicit computation yields

\begin{equation}
\label{starcohe}
| \Xi_{\beta_1} \rangle \star | \Xi_{\beta_2} \rangle= \exp\bigl[
N(\beta_1, \beta_2)\bigr]| \Xi_{\rho_1 \beta_1 - \rho_2 \beta_2} \rangle,
\end{equation}

\noindent
where 
\begin{eqnarray}
\rho_1 &=& -{1 \over 1 + \Phi \cal K} 
\Bigl( H (M^{21})^2+M^{12}(1-HM^{11}) 
\Bigr)={1 \over 2} (1+ MH -C\widetilde M),\nonumber\\
\rho_2&=& {1 \over 1 + \Phi \cal K}\Bigl(H (M^{12})^2+ 
M^{21}(1-HM^{11})\Bigr)={1 \over 2} (1- MH +C\widetilde M),
\end{eqnarray}

\noindent
and 

\begin{eqnarray}
\label{normacohe}
N(\beta_1, \beta_2)& =&{1 \over 2} 
\left( \begin{array}{cc} \beta_1 &
\beta_2 \end{array}\right) \left( \begin{array}{cc} {\cal A} & {\cal B}  \\
{\cal C} &{\cal A} \end{array}\right) 
\left( \begin{array}{c} \beta_1  \\
\beta_2 \end{array}\right)
 \nonumber \\
&=& -{1 \over 2} 
\left( \begin{array}{cc} \beta_1 &
\beta_2 \end{array}\right)  { C \over 2(1-CIH)}\left( \begin{array}{cc} HM
+(M+2)M^{11} & (M+2)M^{12}  \\
(M+2)M^{21} & HM + (M+2)M^{11} \end{array}\right) 
\left( \begin{array}{c} \beta_1  \\
\beta_2 \end{array}\right).
\end{eqnarray}

\noindent
The matrices $\rho_1$ and $\rho_2$ are real symmetric, and they have the 
following properties:

\begin{eqnarray}
& & \rho_1 + \rho_2 =1,\,\,\,\,\,\,\,\,\, \rho_1 \rho_2 =0,\nonumber\\
& & \rho_1^2=\rho_1,\,\,\,\,\,\,\,\,\, \rho_2^2 =\rho_2,
\end{eqnarray}

\noindent 
just as in the bosonic case studied in \cite{RSZhalf}. This means that 
$\rho_1$, $\rho_2$ are orthogonal projectors on complementary subspaces.
We also have
 
$$
C\rho_1 C=\rho_2. 
$$

\noindent 
Notice that the vectors $w^{(\kappa)}$ that we described in the previous
section are eigenvectors of $\rho_{1,2}$. Let us denote by $\sigma_1(\kappa)$ 
and $\sigma_2 (\kappa)$ the corresponding eigenvalues. By using 
(\ref{finaleigen}) and (\ref{hvalue}), we find that

\begin{equation}
\sigma_1 (\kappa) = \begin{cases}
1 ,& \kappa >0, \\
 0 ,& \kappa <0,
\end{cases}
\end{equation}     

\noindent 
with $\sigma_2 (\kappa)=1 -\sigma_1 (\kappa)$. Notice that the eigenvalues 
associated to the vector $\nu$ are $\sigma_1 (\kappa) = \sigma_2 (\kappa) 
= {1\over 2}$. This contradicts in principle the statement that $\rho_1 
\rho_2 =0$, and it gives yet another example of a fact noticed in 
\cite{HM}: formal computations involving inverses of matrices like $1-M^2$ 
become ambiguous when acting on special eigenvectors.

\subsection{Higher--Rank Projectors}

It is obvious from (\ref{starcohe}) that the star multiplication law for 
coherent states becomes particularly simple when $\beta_{1,2}$ are 
eigenvectors of the projectors $\rho_{1,2}$ or combinations thereof. In this 
subsection, we will show that with this choice one finds states that form 
closed subalgebras of the fermionic star algebra. These states can be used 
to obtain new idempotent states that lead to higher--rank projectors, as 
in the bosonic situation \cite{RSZhalf}. The construction is indeed a direct 
generalization of section 5.2 of \cite{RSZhalf}. Let $v$ be an eigenvector 
of $\rho_2$,

\begin{equation}
\rho_1 v =0, \,\,\,\,\,\,\,\,\, \rho_2 v=v,
\end{equation}

\noindent
and define $w=-C v$. Therefore, $\| v\|=\| w\|$, and it follows from 
$C\rho_1 C=\rho_2$ that one will have

\begin{equation}
\label{wdef}
\rho_1 w =w, \,\,\,\,\,\,\,\,\, \rho_2 w=0.
\end{equation}

\noindent
In addition, one has that $v \cdot w= v^T (\rho_1 + \rho_2)w=0$, as in 
\cite{RSZhalf}. Using the explicit expressions 
for $\rho_{1,2}$, one can also show that

\begin{eqnarray}
\label{normavw}
v^T {\cal A}w &=&v^T {\cal C} w = {1 \over 2} v^T M v, \nonumber\\
v^T {\cal B}w &=& {1 \over 2} v^T (1 + C \widetilde M)v,
\end{eqnarray}

\noindent
where the matrices ${\cal A}$, ${\cal B}$ and ${\cal C}$ are the ones 
appearing in (\ref{normacohe}).
 
Consider now the following states, obtained by acting with fermionic 
creation operators on the fermionic sliver $\Xi_f$  (we suppress the brackets 
for notational convenience):

\begin{eqnarray}
\label{flivers}
\Sigma_v &=& \left( {v \over \| v \|}  \cdot \psi^{\dagger} 
\right) \Xi_f , \nonumber\\
\Sigma_w  &=& \left( {w \over \| w\|} \cdot \psi^{\dagger} 
\right) \Xi_f,
\nonumber\\
\Xi_{v,w} &=& \left( { v \over \|v \|} \cdot \psi^{\dagger} \right)
\left( { w \over \| w\|} \cdot \psi^{\dagger} \right)
\Xi_f.
\end{eqnarray}

\noindent
Observe that the state $\Xi_{v,w}$ is Grassmann even, since fermions only 
appear via bilinears, while the $\Sigma_{v,w}$ states are Grassmann odd. 
Consider now the coherent states $\Xi_{\beta_1}$, $\Xi_{\beta_2}$, where 
$\beta_1 = \theta_1 v + \theta_2 w$, $\beta_2 = \hat \theta_1 v + \hat 
\theta_2 w$ and $\theta_{1,2}$, $\hat \theta_{1,2}$ are arbitrary Grassmann 
variables. It is simple to show, by computing the star product 
$\Xi_{\beta_1}\star \Xi_{\beta_2}$, that the states defined in 
(\ref{flivers}) satisfy the following subalgebra of the star product, in 
the fermionic matter sector:

\begin{align}
\label{fliversub}
\Xi_f \star \Sigma_v &=0, 
&\Xi_f \star \Sigma_w &= -\Sigma_w, \nonumber 
\\ 
\Sigma_v \star \Xi_f &=\Sigma_v,  
&\Sigma_w \star \Xi_f &= 0, \nonumber\\
\Sigma_v \star \Sigma_v &=0, 
&\Sigma_w \star \Sigma_w &=0,\nonumber\\
\Sigma_v \star \Sigma_w &={\cal A}_v \Xi_f - \Xi_{v,w}, 
& \Sigma_w \star \Sigma_v&= -{\cal B}_v \Xi_f,\nonumber\\
\Xi_f \star \Xi_{v,w}&={\cal A}_v\Xi_f, 
&\Xi_{v,w}\star \Xi_f &= {\cal A}_v\Xi_f,\nonumber\\
\Sigma_v \star \Xi_{v,w}&={\cal A}_v\Sigma_v, 
 &\Sigma_w \star \Xi_{v,w}&={\cal B}_v \Sigma_w, \nonumber\\
\Xi_{v,w}\star \Sigma_v & =  -{\cal B}_v\Sigma_w , 
&\Xi_{v,w}\star \Sigma_w & =-{\cal A}_v \Sigma_w,
\end{align}
and finally
\begin{equation}
\Xi_{v,w}\star \Xi_{v,w}= {\cal A}_v ({\cal A}_v-{\cal B}_v) \Xi_f + 
{\cal B}_v \Xi_{v,w}.
\end{equation}

\noindent
In these equations we have introduced the notation 

\begin{equation}
{\cal A}_v ={v^T {\cal A} w \over \| v \|^2}, \,\,\,\,\,\,\,\,\, 
{\cal B}_v ={ v^T {\cal B}w \over \| v \|^2}. 
\end{equation}

\noindent  
One can also find the BPZ norm of these states by computing 
$\langle \Xi_{\beta} | \Xi_{\beta} \rangle$, with $\beta =\theta_1 v +
\theta_2 w$. In this computation one has to evaluate the inner products

\begin{eqnarray}
v^T {F \over 1-F^2}w&=&-\| v\|^2 {\cal A}_v, \nonumber\\
v^T {1 \over 1-F^2}v&=&w^T {1 \over 1-F^2}w =\| v\|^2 {\cal B}_v,
\end{eqnarray}

\noindent
as it can be checked by using the explicit expression for $F$ and 
the fact that $v$, $w$ are eigenvectors of $\rho_{1,2}$. One obtains 
in the end: 

\begin{eqnarray}
 \langle \Sigma_v | \Sigma_v\rangle &=& {\cal B}_v 
\langle \Xi_f | \Xi_f \rangle, \nonumber\\ 
\langle \Sigma_v | \Sigma_w \rangle&=& 0,\nonumber\\
 \langle \Xi_f | \Xi_{v,w}\rangle &=&-{\cal A}_v \langle \Xi_f | \Xi_f
\rangle,\nonumber\\ 
\langle \Xi_{v,w} | \Xi_{v,w}\rangle &=&-({\cal A}_v^2 +{\cal B}_v^2)
\langle \Xi_f | \Xi_f \rangle,
\end{eqnarray}

\noindent
together with their BPZ conjugates (notice that that BPZ conjugation 
exchanges $v \leftrightarrow w$).

One can now use this subalgebra in order to generate new solutions 
to the idempotency condition, and thus new solutions to the vacuum 
superstring field theory equations of motion. This one can do by taking 
the most general combination of the four states, $\Xi_{f}$, $\Xi_{v,w}$, 
$\Sigma_{v}$ and $\Sigma_{w}$, (and with the appropriate Chan--Paton factors, 
since the $\Sigma$'s are Grassmann odd). One finds in this way two types of 
new solutions:

\bigskip

\noindent
\textbf{1)} There is one new solution, which is Grassmann even. It is 
given by

\begin{equation}
\chi_f=
\alpha \Xi_f + \beta \Xi_{v,w},
\end{equation}

\noindent
where $\chi_{f} \star \chi_{f} = \chi_{f}$ provided one chooses

\begin{equation}
\alpha = -{ {\cal A}_v \over {\cal B}_v}, \,\,\,\,\,\,\,\,\, 
\beta={1 \over {\cal B}_v}.
\end{equation}

\noindent
One has that

\begin{equation}
\chi_f \star \Xi_f =\Xi_f \star \chi_f =0,
\end{equation}

\noindent
and also 

\begin{equation}
\langle \chi_f | \chi_f \rangle = \langle \Xi_f | \Xi_f \rangle, 
\,\,\,\,\,\,\,\,\, \langle \chi_f | \Xi_f \rangle =0.
\end{equation}

\noindent
Therefore, we see that if one interprets the fermionic sliver as a projector 
in the space of string fields, the string field $\chi_f$ is a projector on 
an orthogonal subspace and their sum is then a higher rank projector, as in 
\cite{RSZhalf}. Indeed, the fermionic sliver is a rank--one projector on the 
fermionic sector of the space of half--string functionals. The best way to 
see this would be of course to construct a half--string formalism for the 
fermion fields. Unfortunately we have not been able to do that, as we have 
not found good boundary conditions for the split fermions. However, one 
can still bosonize the fermions and reduce the problem to the case already 
analyzed in \cite{RSZhalf, GT, GT2}. In fact, bosonization was used in 
\cite{GT2} to show that the ghost part of the bosonic sliver is also 
a rank--one projector on the ghost sector of the space of 
half--functionals\footnote{For a discussion on the bosonization of the 
interaction vertex of the superstring in the operator formalism, see 
\cite{samuel}.}.

\bigskip

\noindent
\textbf{2)} There are two families of new solutions, which have both a 
Grassmann even and a Grassmann odd piece. The first one is

\begin{equation}
\label{fsolone}
\Xi_{\ell}=\Xi_f \otimes \mathbf{1} + \ell\ \Sigma_v \otimes \sigma_{1}, 
\,\,\,\,\,\,\,\,\, \ell \in {\bf R},
\end{equation}

\noindent
while the second one is 

\begin{equation}
\label{fsoltwo}
\chi_{\ell}=\chi_f \otimes \mathbf{1} + \ell\ \Sigma_w \otimes \sigma_{1}, 
\,\,\,\,\,\,\,\,\, \ell \in {\bf R}.
\end{equation}

\noindent
These string fields are idempotent for arbitrary real $\ell$, since the 
Grassmann odd piece is a nilpotent state, and they have the same norm 
for any $\ell$, which is the norm of the fermionic sliver. Moreover, one
can show that $\Xi_{\ell}$ and $\chi_{\ell}$ are related to $\Xi$ and 
$\chi$ by gauge transformations at the vacuum. Indeed, using nilpotency of 
$\Sigma_{v,w}$ and the fact that

\begin{equation}
[\Sigma_v, \Xi_f]_\star= \Xi_f, \,\,\,\,\,\,\,\,\,
[\Sigma_w, \chi_f]_\star= \chi_f,
\end{equation}

\noindent
one finds

\begin{eqnarray}
\Xi_{\ell}&=&{\rm e}^{\ell\ \Sigma_v \otimes \sigma_1} \star \Xi_f \star
{\rm e}^{-\ell\ \Sigma_v \otimes \sigma_1},\nonumber \\
\chi_{\ell}&=&{\rm e}^{\ell\ \Sigma_w \otimes \sigma_1} \star \chi_f \star
{\rm e}^{-\ell\ \Sigma_w \otimes \sigma_1},
\end{eqnarray}

\noindent
{\it i.e.}, in the terminology of \cite{RSZhalf} $\Xi_{\ell}$ and 
$\chi_{\ell}$ are star rotations of $\Xi_f$ and $\chi_f$. But on the 
other hand, star rotations are indeed gauge transformations at the 
vacuum as it follows from (\ref{gaugem}). In the case we are 
considering, the gauge parameter is simply given by $\Xi_m= \ell\ 
\Sigma_{v,w} \otimes \sigma_1$. 

In order to construct higher--rank projectors, we have used simultaneous 
eigenvectors of the projectors $\rho_1$ and $\rho_2$. These are precisely 
the $w^{(\kappa)}$ that we have found in section 5, if one assumes that all 
the eigenvectors of these matrices are the eigenvectors of ${\rm K}_1$. In 
this case, one can take for $v$ any vector $w^{(-\kappa)}$ with $\kappa>0$, 
and then $w=w^{(\kappa)}$. The states defined in (\ref{flivers}) give a 
family of fermionic subalgebras parametrized by $\kappa >0$, with 
coefficients

\begin{equation}
{\cal A}_v = -{{\rm e}^{-\kappa \pi/2} \over 
1+{\rm e}^{-\kappa \pi/2}}, \,\,\,\,\,\,\,\,\, 
{\cal B}_v={1 \over 1+{\rm e}^{-\kappa \pi/2}}. 
\end{equation}

\noindent 
Notice that we have normalized these states by introducing a factor 
$1/\|v \|$. In this way, the norms of $v,w$ do not appear in the star 
subalgebra nor in the BPZ products. Since the vectors $w^{(\kappa)}$ have 
infinite norm, this normalization factor actually vanishes. Observe, 
however, that the norm of the (super)sliver is also strictly zero since it 
contains a positive power of $({\rm det}(1+X))$, and the matrix $X$ is known 
to have an eigenvalue $-1/3$ \cite{HM, MT, RSZspectro}. In that respect, the 
states we have constructed are not essentially different. We should add that 
the same thing happens to the higher rank projectors constructed in 
\cite{RSZhalf}: they are constructed from eigenvectors of the bosonic 
projectors, which have infinite norm \cite{MT, RSZspectro}, and the 
construction involves dividing by this norm. This is yet another 
manifestation of the rather singular structure of the idempotents of the 
string field star algebra.

Some remarks are now in order. It is simple to see from (\ref{starcohe})
and the fermionic subalgebra (\ref{fliversub}), that associativity of the
star product does not hold in the fermionic sector. The breakdown of
associativity is however rather mild, as it holds up to signs. It is known
that in order to have associativity of the string star product both the
three vertex and the four vertex need to be cyclic (see for example
\cite{LPP}). Although the three vertex analyzed in section 3 is indeed
cyclic, it has been shown by Bogojevi\' c, Jevicki and Meng \cite{jevicki}
that in the fermionic matter sector the four vertex is not
cyclic. Cyclicity is however expected to hold once we restrict ourselves to
the GSO(+) sector, and this is in agreement with the algebra
(\ref{fliversub}).

\section{The Geometric Supersliver and the (Super)Ghost Sector}

So far, we have restricted ourselves to the matter sector. In order to have 
a complete picture, we still have to make a proposal for the the vacuum BRST 
${\cal Q}$ operator, and one has to solve the equations of motion in the 
ghost sector (\ref{ghosteq}). In this section we will show that the 
ghost/superghost part of the geometric supersliver satisfies (\ref{ghosteq}) if 
we take  ${\cal Q}$ to be the canonical BRST operator recently proposed by 
Gaiotto, Rastelli, Sen and Zwiebach \cite{GRSZ} for the bosonic string, and 
that we shall denote in the following by ${\cal Q}_{\rm GRSZ}$. Observe that 
this implies that the full geometric supersliver is a solution to the full 
superstring field theory equations of motion. Therefore, a natural proposal 
for vacuum superstring field theory is to take 
${\cal Q}={\cal Q}_{\rm GRSZ}$, and postulate that the maximal $D9$--brane is 
described by the full geometric supersliver. 

Notice that the string field in 
Berkovits' theory has ghost and picture number zero, and therefore the 
geometric supersliver is a good string field. This is in contrast to bosonic 
string theory, where the string field has ghost number one and therefore the 
sliver is not an acceptable string field. Indeed, the $D25$--brane is 
conjecturally described by the twisted sliver, whose algebraic construction 
was presented in \cite{hata} and has later been constructed in BCFT in 
\cite{GRSZ}. The twisted sliver has in fact ghost number one, as required 
by cubic bosonic string field theory. 

Let us then analyze the equation (\ref{ghosteq}). We have seen in section 2 
that idempotency 
of the string field seems to be even more useful in superstring field theory, 
where it reduces drastically the nonlinearity of the equation of motion. In 
fact, it is easy to see that an idempotent ghost/superghost state 
satisfying $\Phi_{g} \star \Phi_{g} = \Phi_{g}$ reduces the 
WZW equation of motion (\ref{ghosteq}) to a simpler form. If $\Phi_g$ is 
idempotent, the exponential is linearized as

\begin{equation}
e^{\Phi_{g}} = {\cal I} + \left( e - 1 \right) \Phi_{g},
\end{equation}

\noindent
and so the equation of motion becomes

\begin{equation}
\eta_0 \left\{ \left( {\cal I} + \left( \frac{1}{e} - 1 \right) \Phi_{g} 
\right) {\cal Q} \Phi_{g} \right\} =0.
\end{equation}

\noindent
It is clear that this equation of motion is solved if

\begin{equation}
{\cal Q} \Phi_{g} =0.
\end{equation}

Let us then assume that the vacuum BRST operator is the one chosen 
recently by Gaiotto, Rastelli, Sen and Zwiebach \cite{GRSZ} for the 
bosonic string,

\begin{equation}
{\cal Q_{\mathrm{GRSZ}}} = \frac{1}{2i} \left( c(i) - c(-i) \right).
\end{equation}

\noindent
In terms of oscillators, this operator is given by

\begin{equation}
{\cal Q_{\mathrm{GRSZ}}} = \sum_{n=0}^{\infty} (-1)^n {\cal C}_{2n},
\end{equation}

\noindent
where 

\begin{eqnarray}
{\cal C}_n &=& c_n + (-1)^n c_n, \,\,\,\,\,\,\,\,\, n \not=0,
\nonumber\\
{\cal C}_0 &=& c_0.
\end{eqnarray}

\noindent
We can now show that the ghost part of the supersliver is annihilated by 
${\cal Q_{\mathrm{GRSZ}}}$, and therefore solves the equations of 
motion\footnote{We thank L.~Rastelli for pointing this out to us.}. First of 
all, notice that the ghost part of the supersliver factorizes into $bc$ and 
$\beta\gamma$ pieces, 

\begin{equation}
|\Xi_g\rangle =|\Xi_{bc} \rangle \otimes |\Xi_{\beta\gamma} \rangle.
\end{equation}

\noindent
Since the choice ${\cal Q_{\mathrm{GRSZ}}}$ does not involve the superghosts, 
it is enough to show that 

\begin{equation}
\label{magic}
{\cal Q_{\mathrm{GRSZ}}}|\Xi_{bc}\rangle=0.
\end{equation}

\noindent
We shall do this in two distinct ways. First, we use a geometric argument akin 
to that in \cite{GRSZ}. Secondly, we shall prove it by using oscillator 
methods. 

The geometric argument goes as follows. The supersliver is defined by the 
following relation, 

\begin{equation}
\langle \Xi | \phi \rangle = \langle f \circ \phi \rangle,
\end{equation}

\noindent
where $f(z) = \arctan (z)$, and $|\phi\rangle$ is any Fock state. If 
one now acts with the arbitrary Fock state $\langle \phi|$ on (\ref{magic}), 
one finds

\begin{equation}
 \langle \phi\ {\cal Q_{\mathrm{GRSZ}}} | \Xi \rangle = {1\over 2i} 
\langle f \circ \phi(0) \Bigl((f'(i))^{-1} c(i \infty) - (f'(-i))^{-1} c(-i 
\infty) \Bigr) \rangle.
\end{equation}

\noindent
But $(f'(\pm i ))^{-1}=0$, and therefore the above correlator is zero.

Let us now give an oscillator proof. Using the methods of \cite{LPP}, 
it is not too hard to show that the $bc$ part of the (super)sliver is given 
by a squeezed state of the form

\begin{equation}
|\Xi_{bc} \rangle = \exp \Bigl( \sum_{s,i} c_{-s} S_{si} b_{-i} \Bigr)
|0 \rangle,
\end{equation}

\noindent
where $s = -1, 0, 1, \ldots$, $i = 2, 3, \ldots$, and $S_{si}$ is given by 
the double residue:

\begin{equation}
S_{si} = \oint_0 {dz \over 2\pi i} {1 \over z^{s-1}} \oint_0 {dw \over 2\pi i} 
{1 \over w^{i+2}} {(f'(z))^2 (f'(w))^{-1} \over f(z)-f(w)} \biggl( {f(w) \over 
f(z)} \biggr)^3.
\end{equation} 

\noindent
A different expression for this matrix has been given in \cite{david}. If we 
now define $U = \sum_{s,i} c_{-s}S_{si}b_{-i}$, one has that for $n \ge 2$,

\begin{equation}
c_n U = U c_n -\sum_s c_{-s}S_{si}.
\end{equation}

\noindent
Using this result one can easily show that (\ref{magic}) holds if and only if 
the matrix $S$ satisfies

\begin{equation}
\label{eigenghost}
\sum_{n=1}^{\infty} S_{2k, 2n} (-1)^n = (-1)^k,
\end{equation}

\noindent
where we have also used that, due to twist invariance, $S_{si}=0$ if $s + i = 
{\rm odd}$. The above equation says essentially that $S$ has an eigenvector 
with eigenvalue $1$. One can check that (\ref{eigenghost}) is true by using the 
explicit representation of $S$ as a double residue and the techniques of 
\cite{RSZspectro}. Indeed, since 

\begin{equation}
\sum_{n=1}^{\infty} (-1)^n w^{-2n-2 }= { w^{-2} \over 1 + w^2},
\end{equation}

\noindent
we have to deform the $w$ contour to pick the residue at $w=z$, and this 
yields

\begin{equation}
\label{eigenghost2}
\sum_{n=1}^{\infty} S_{2k,2n} (-1)^n =\oint_0 {dz \over 2\pi i } {1 \over 
z^{2k-1}} {1 \over 1+z^2} = (-1)^k ,
\end{equation}

\noindent
as we wanted to show. This gives yet another proof of (\ref{magic}), and also 
establishes a property of $S$ that may be relevant in future investigations. 

Notice that in order to annihilate the identity the BRST operator of 
\cite{GRSZ} has to be regularized in an appropriate way. It is also
immediate to observe that this regularization does not affect the 
above computations.

\section{Conclusions and Future Directions}

In this paper we have taken the first steps towards the construction of vacuum
superstring field theory. More concretely, we have shown that idempotent
states play a distinctive role in Berkovits' string field theory, and after
clarifying the structure of the fermionic vertex in the NS sector we have
given an explicit algebraic construction of the fermionic sliver. We have
also explored some aspects of the star algebra. In particular, we have 
computed the spectrum of the fermionic Neumann matrices and we have
constructed higher--rank projectors by using closed star subalgebras 
obtained from coherent states. Finally, we have shown that the 
geometric sliver is a solution to the superstring field theory 
equations of motion including both matter and ghost sectors.

Clearly, many things remain to be done. There are some obvious open problems
that one should address to put this construction on a firmer ground, 
which we now list as directions for future research.

\bigskip

\noindent
$\bullet$ The most pressing problem is to construct solutions describing 
the various BPS and non--BPS $D$--branes of Type IIA superstring theory. 
It is natural to conjecture that the supersliver describes the tachyonic 
vacuum of the $D9$--brane, but a necessary test is to verify that one 
can construct other $D$--branes with the right ratio of tensions. In 
\cite{RSZclassical}, lower dimensional $D$--branes were constructed in the 
bosonic case by first identifying the sliver state with the maximal 
$D25$--brane and then exploiting the spacetime dependence of the vertex. A 
more general construction was later described in \cite{RSZboundary} and 
implemented in detail in \cite{Mukho}. It should be possible to adapt this 
construction to the supersymmetric case, although there might be some subtle 
points that need to be addressed. For example, it is not obvious to us 
how one would reproduce the mod two behavior of the $D$--brane descent 
relations in the superstring case, {\it i.e.}, the fact that in the IIA theory 
$Dp$--branes with odd $p$ are unstable while $Dp$--branes with even $p$ are 
stable, and in particular the fact that unstable and stable $D$--brane tensions 
differ with an extra ${\sqrt 2}$ factor. One possibility is that this question 
of ``BPS versus non--BPS'' brane solutions could also be associated to the 
construction of solutions to vacuum superstring field theory only in the 
GSO$(+)$ sector or in both the GSO$(\pm)$ sectors. Another possibility 
may have to do with the introduction of the Grassmann odd state 
$G_{-1/2} | \Xi \rangle$ in the game. But surely the most straightforward 
way to proceed would be to follow the methods of \cite{RSZboundary, Mukho}.

\bigskip

\noindent
$\bullet$ One should also understand the structure of the ghost and superghost 
components of the sliver. Notice that in Berkovits' theory the correlation 
functions that enter into the star product are defined in the large Hilbert 
space and, therefore, one should have a construction of the superghost vertex 
in terms of the bosonized superconformal ghosts. The full analysis of the 
ghost/superghost sector will be probably necessary in order to identify the 
closed superstrings at the nonperturbative vacuum, perhaps along the lines of 
\cite{GRSZ}.

\bigskip

\noindent
$\bullet$ It would be interesting to develop a half--string formalism
\cite{BCNT} in the fermionic sector. This would make clear some of the 
properties of the fermionic sliver, like the fact that it is a rank--one
projector. As we pointed out in section 6, a way to see this is to bosonize
the fermions, but it would be more convenient to have an explicit
representation in terms of fermionic oscillators. 

\bigskip

\noindent
$\bullet$ Although in this paper we have focused on Berkovits' superstring
field theory, there exists another proposal for superstring field theory
of the NS sector, which is cubic and has been also used to test Sen's 
conjectures (see, \textit{e.g.}, the recent review \cite{arefeva}). In 
this cubic superstring field theory, where the string field has picture 
number zero and ghost number one, one can immediately extend all of the 
results of bosonic VSFT: assuming a pure ghost/superghost BRST operator, and
factorization of the solution, the equation of motion in the matter 
part again reduces to idempotency of the string field. Since the star product 
is kept the same, all the results of this paper, concerning the 
fermionic matter sector, are as well valid for the cubic superstring field 
theory. The ghost sector, however, will probably require some sort 
of twisted ghost sliver as in \cite{GRSZ}.

\bigskip

\noindent
\textsf{\textbf{Added note:}} After this paper was completed, the paper
\cite{nsmattersliver} appeared, which has some overlap with sections 3 and
4 in this paper and constructs the fermionic sliver in the context of the
cubic superstring field theory.

\section*{Acknowledgements}
We would like to thank N.~Berkovits, A.~Jevicki and specially L.~Rastelli for very 
useful discussions and correspondence. MM is supported by the grant 
NSF--PHY/98--02709. RS is supported in part by funds provided by the 
Funda\c c\~ao para a Ci\^encia e Tecnologia, under the grant Praxis XXI 
BPD--17225/98 (Portugal).

\vfill

\eject

\appendix

\section{Appendix}

In this appendix, we show that the explicit expressions given in 
(\ref{expli}) and (\ref{Ks}) satisfy equations (\ref{idenm}) and (\ref{Ue}), 
respectively. Since this is very similar to the bosonic case analyzed in 
\cite{GJ}, we shall only give a few details. In the case of the identity, 
we are required to prove that $(1-M) I ={\widetilde M}$. The only thing 
we actually need is the following result:

\begin{equation}
\label{sumbin}
\sum_{\ell=0}^{\infty} {(-1)^{\ell} \over \ell+a} \left( \begin{array}{c} 
-1/2 \\ \ell\end{array}\right)= {\Gamma({1\over 2})\Gamma (a)
\over \Gamma(a+{1 \over 2})}.
\end{equation}

\noindent
From this, one deduces that

\begin{equation}
\label{sumu}
\sum_{\ell=0}^{\infty}{\hat u_{2\ell}\over 2\ell + a + 1} =\begin{cases}
{\pi \over 2} \hat u_a,& a \, {\rm even},\\
{1 \over a \hat u_a}, & a \, {\rm odd},\end{cases}
\end{equation}

\noindent
and this is enough to prove (\ref{idenm}). For the interaction vertex, 
one has to prove the following equations:

\begin{eqnarray}
\label{overtwo}
(M+2)(U + \overline U)&=&-2 \widetilde M -2(U + \overline U),\nonumber\\
(M+2)(U-\overline U)&=&2 {\sqrt 3}iC,
\end{eqnarray}

\noindent
where the matrices $U+\overline U$, $U-\overline U$ are given in (\ref{uus}).
The necessary ingredients to prove (\ref{overtwo}) are the following.
First, one can show that the coefficients $g_n$ defined in (\ref{gseries})
satisfy the recursion relation:

\begin{equation}
\label{recurone}
{1 \over 3} g_n =(n+1) g_{n+1}-(n-1) g_{n-1}.
\end{equation}

\noindent
Next, define as in \cite{GJ} the following sums:

\begin{eqnarray}
\label{sums}
O_n &=&\sum_{m=2\ell+1} {g_m \over n+m },\nonumber\\
E_n &=&\sum_{m=2\ell} {g_m \over n+m}.
\end{eqnarray}

\noindent
These sums can be written as integrals,

\begin{eqnarray}
\label{ints}
O_n &=&{1 \over 2}\int_1^{\infty} {dt\over t^{n+1}}
 \Biggl[ \biggl( {t+1 \over t-1}\biggr)^{1/6} -
\biggl( {t+1 \over t-1}\biggr)^{1/6}\Biggr],\nonumber\\
E_n &=&{1 \over 2} \int_1^{\infty} {dt\over t^{n+1}}
\Biggl[ \biggl( {t+1 \over t-1}\biggr)^{1/6} +
\biggl( {t+1 \over t-1}\biggr)^{1/6}\Biggr],
\end{eqnarray}

\noindent
and using this representation one can show that they satisfy the recursion 
relations:

\begin{eqnarray}
\label{maisrec}
(n+1)E_{n+1}&=&{1 \over 3} O_n + (n-1)E_n ,
\nonumber\\
(n+1)O_{n+1}&=& {1 \over 3} E_n + (n-1)O_n.
\end{eqnarray}

\noindent
To evaluate these sums, we proceed as in \cite{GJ}. On the one hand,
we have

\begin{equation}
g_{2\ell}={1\over 2\pi i } \oint {dz \over z^{2\ell+1}}{1 \over 2}
\Biggl[ \biggl( {z+1 \over z-1}\biggr)^{1/6} -
\biggl( {z+1 \over z-1}\biggr)^{1/6}\Biggr],
\end{equation}

\noindent
where the contour is around the origin. On the other hand, when $\ell$ is 
greater than zero, one can deform the contour in the above integral to the 
real axis and obtain

\begin{equation}
O_{2\ell} = \pi g_{2\ell}, \quad \ell\ge 1.
\end{equation}

\noindent
Similarly, one proves that

\begin{equation}
E_{2\ell +1} = \pi g_{2\ell+1}, \quad \ell\ge 0.
\end{equation}

\noindent
The value of $O_0$ can be evaluated by direct integration. After
performing the change of variables $x=\tanh ((\log t)/2)$, one finds

\begin{equation}
O_0=\int_0^{1} {dx\over 1-x^2} \bigl( x^{-1/6} - x^{1/6}\bigr)=
{1\over 2}\Bigl(\psi \Bigl({7\over 12}\Bigr)-\psi \Bigl({5\over 12}
\Bigr) \Bigr).
\end{equation}

\noindent
We then find

\begin{equation}
\label{ceroval}
O_0=\pi -{{\sqrt 3}\over
2}.
\end{equation}

\noindent
Using this value and the recursion relations, one can obtain $O_{-2\ell}$,
$E_{-2\ell-1}$ as well. To evaluate the other sums, we follow the procedure 
in the Appendix of \cite{GJ}. First, define the series

\begin{equation}
S_n =\begin{cases} E_n,& n=2k,\\
O_n, & n=2k+1.\end{cases}
\end{equation}

\noindent
Since the sums satisfy the recursion relation (\ref{maisrec}), $S_n$
satisfies the recursion relation of the coefficients $g_n$,
(\ref{recurone}). There is another solution to this relation which is given
by

\begin{equation}
S_n = 3 S_1 g_n + 3 \sum_{m=0}^{n-1} (-1)^m {g_m g_{n-m} \over m+1}.
\end{equation}

\noindent
To derive this, one first writes a differential equation for the
function $S(x)=\sum_{n=1}^{\infty} S_n x^n$ by using the recursion
relation. The details are exactly like the ones in \cite{GJ}. To have the
complete solution to the problem, we then just have to evaluate $S_1$,

\begin{equation}
S_1=1 + {1\over {\sqrt 3}}\log \biggl( { {\sqrt 3}-1
\over {\sqrt 3}+1}\biggr).
\end{equation}

\noindent
Notice that $S_m \sim 1/m$, and one has $mS_m \rightarrow 1$ as
$m\rightarrow 0$. The recursion relation also implies that $S_{-m}$
diverges for $m=-1, -2, \cdots$, but $mS_{-m-n}$ with $n>0$ has a finite 
limit as $m$ goes to zero that can be evaluated using the recursion relations.

With these ingredients, we can already prove very easily the first equation in
(\ref{overtwo}). For example, in this proof one has to evaluate the
quantity

\begin{equation}
A_m =(-1)^m m(m+1)\bigl( g_{m+1}S_m -S_{m+1}g_m \bigr).
\end{equation}

\noindent
Using the recursion relations, one can see that $A_m$ does not depend on
$m$, therefore $A_m =A_1=1/3$. In order to prove the second equation in 
(\ref{overtwo}), one needs some extra ingredients to deal with the diagonal 
terms, since these involve the sums

\begin{eqnarray}
\widetilde O_n& =&\sum_{m=2\ell+1} {g_m \over (n+m)^2 },\nonumber\\
\widetilde E_n &=&\sum_{m=2\ell} {g_m \over (n+m)^2}.
\end{eqnarray}

\noindent
These sums have the integral representation

\begin{eqnarray}
\label{intstwo}
\widetilde O_n &=&{1 \over 2} \int_1^{\infty}dt {\log \,t \over t^{n+1}}
 \Biggl[ \biggl( {t+1 \over t-1}\biggr)^{1/6} -
\biggl( {t+1 \over t-1}\biggr)^{1/6}\Biggr],\nonumber\\
\widetilde E_n &=& {1 \over 2} \int_1^{\infty} dt {\log \, t\over t^{n+1}}
 \Biggl[ \biggl( {t+1 \over t-1}\biggr)^{1/6} +
\biggl( {t+1 \over t-1}\biggr)^{1/6}\Biggr],
\end{eqnarray}

\noindent
and using them one can prove the recursion relations:

\begin{eqnarray}
\label{maisrecs}
(n+1)\widetilde E_{n+1}&=&{1 \over 3} \widetilde O_n + (n-1)\widetilde E_n
+E_{n+1}-E_{n-1},
\nonumber\\
(n+1)\widetilde O_{n+1}&=& {1 \over 3} \widetilde E_n + (n-1)\widetilde O_n
+ O_{n+1}-O_{n-1}.
\end{eqnarray}

\noindent
These can be solved as in \cite{GJ}, but we shall not need their explicit
expression, and in the evaluation of the relevant quantities
it suffices to use the recursion relations they satisfy. For example, in 
the proof of the second equation in (\ref{overtwo}) one has to compute

\begin{equation}
C_m = m(m+1)\bigl( g_{m+1}\widetilde S_m - \widetilde
S_{m+1}g_m \bigr),
\end{equation}

\noindent
where $\widetilde S_n$ is defined as follows:

\begin{equation}
\widetilde S_n =\begin{cases} \widetilde O_n,& n=2k,\\
\widetilde E_n, & n=2k+1.\end{cases}
\end{equation}

\noindent
Using the recursion relations (\ref{recurone}) and (\ref{maisrecs}), as
well as (\ref{ceroval}), one can show that

\begin{equation}
C_m = {\pi\over 3} \sum_{l=0}^m (-1)^l g_{m-l}^2 -\pi  g_m g_{m+1} -{ \pi
{\sqrt 3} \over 6}.
\end{equation}

\noindent
Taking into account these results, the proof of the second equation
in (\ref{overtwo}) is immediate.


\vfill

\eject

\bibliographystyle{plain}

\end{document}